\documentclass[useAMS,usenatbib]{mn2e}

\usepackage{graphicx}
\newcommand{\feh}{\ensuremath{\rm [Fe/H]}}
\newcommand{\msun}{\ensuremath{{\rm M}_{\odot}}}
\newcommand{\zsun}{\ensuremath{{\rm Z}_{\odot}}}
\usepackage{xspace}
\newcommand{\galev}{\textsc{galev}\xspace}

\usepackage[usenames]{color}
\usepackage[normalem]{ulem}

\title[GALEV: code, input physics and web-interface]{GALEV evolutionary synthesis models\\ 
  I. Code, input physics and web-interface}

 \author[The GALEV team]{Ralf
  Kotulla$^{1}$\thanks{E-mail: \mbox{r.kotulla@herts.ac.uk},
    \mbox{u.fritze@herts.ac.uk}, \mbox{pweilbacher@aip.de},
    \mbox{p.anders@astro.uu.nl}}, Uta Fritze$^{1}$, Peter
  Weilbacher$^{2}$, and Peter Anders$^{3}$: the GALEV team\\
  $^{1}$ Centre for Astrophyics Research, University of Hertfordshire, College
  Lane, Hatfield AL10 9AB, United Kingdom\\
  $^{2}$ Astrophysikalisches Institut Potsdam, An der Sternwarte 16, 14482
  Postdam, Germany\\
  $^{3}$ Sterrekundig Instituut, Princetonplein 5, 3584 CC Utrecht, The Netherlands}
\begin{document}

\date{}

\pagerange{\pageref{firstpage}--\pageref{lastpage}} \pubyear{2008}

\maketitle

\label{firstpage}

\begin{abstract}
\galev evolutionary synthesis models describe the evolution of stellar
populations in general, of star clusters as well as of galaxies, both in terms
of resolved stellar populations and of integrated light properties over
cosmological timescales of $\geq 13$ Gyr from the onset of star formation
shortly after the Big Bang until today.

For galaxies, \galev includes a simultaneous treatment of the chemical
evolution of the gas and the spectral evolution of the stellar content,
allowing for what we call a chemically consistent treatment: We use input
physics (stellar evolutionary tracks, stellar yields and model atmospheres)
for a large range of metallicities and consistently account for the increasing
initial abundances of successive stellar generations.

Here we present the latest version of the \galev evolutionary synthesis models
that are now interactively available at \textbf{http://www.galev.org}. We review
the currently used input physics, and also give details on how this physics is
implemented in practice. We explain how to use the interactive web-interface to
generate models for user-defined parameters and also give a range of
applications that can be studied using \galev, ranging from star clusters,
undisturbed galaxies of various types E $\ldots$ Sd to starburst- and dwarf
galaxies, both in the local and the high-redshift universe.
\end{abstract}


\section{Introduction}
\label{sec:intro}
\galev (short for GALaxy EVolution) evolutionary synthesis models have been
developed over many years. They were published in several steps and under a
variety of first author names, reflecting the number of students who have
contributed their respective shares to the development. \galev models include
the spectral evolutionary synthesis of a stellar population with arbitrary
star formation history on the basis of the time evolution of the stellar
population in the Hertzsprung-Russell diagram, as well as a detailed chemical
evolution model for the ISM in terms of a large number of individual element
abundances. \galev models have a wide range of application from star clusters
(SCs) to resolved nearby galaxies, to more distant galaxies observed in terms of
integrated spectra and photometry, all through galaxies at high redshifts.

Previous applications cover the range from star clusters, normal galaxies E,
$\ldots$ Sd, dwarf galaxies with and without starbursts, tidal dwarf galaxies,
interacting and merging galaxies with their major starbursts, galaxy
transformation processes in galaxy clusters, high redshift galaxies with and
without starbursts and post-starbursts and damped Lyman-$\alpha$ absorbers. An
early attempt at coupling \galev evolutionary synthesis models into a
cosmodynamical structure formation simulation was presented in
\cite{Contardo+98}.

\galev models are now widely used throughout the community. To facilitate
access to the latest developments we here present a user-friendly
and customized web-interface. It allows access to already available models for
the evolution of star clusters of various metallicities, and galaxies of all
types both in terms of their time evolution for comparison with observations
in the Local Universe and in terms of their redshift evolution. Furthermore it
allows the user to run new models for specific applications.

The philosophy for \galev models is to keep them simple with as small a number of
free parameters as possible, and have them predict a large range of
observational properties, which -- in comparison with observations -- constrain
the few free parameters. At the present stage, \galev models are 1-zone models
without spatial resolution and without any dynamics included.  Future prospects
are the consistent inclusion of dust detailing absorption and reemission as a
function of gas content, metallicity and galaxy type, and the coupling with a
dynamical model for stars and gas, including a star formation criterion and an
appropriate feedback description, to cope with spatially resolved galaxy data.

\section{The GALEV code: An overview}
\label{sec:code}
\subsection{Evolutionary synthesis for star clusters and galaxies}
\label{subsec:code:evsyn}
Our \galev evolutionary synthesis models have many properties in common with
the evolutionary synthesis codes from other groups, e.g. BC03
\citep{BruzualCharlot03}, PEGASE \citep{FiocRocca97} and Starburst99
\citep{Leitherer+99}, just to name a few, in that all these codes trace the
evolution of the stellar population in terms of integrated spectra and/or
colours for simple and composite stellar populations.

In contrast to evolutionary synthesis, \emph{stellar population synthesis}
\citep[e.g.][]{Oconnell76,Oconnell80} or \emph{differential synthesis}
  \citep[e.g.][]{Pickles85a,Pickles85b,Pickles+85} attempts to find the best
linear combination of stellar spectra from some library to fit an observed
galaxy spectrum . This approach usually achieves very good fits but is limited
to a \emph{status quo} description and has difficulties to prove the uniqueness
of its solutions. The existence of a stellar Initial Mass Function (IMF) and
some continuous Star Formation History (SFH) can be imposed as boundary
conditions via Lagrangian multipliers.  The major advantages of the population
synthesis approach are that it can give valuable first guesses for unknown SFHs
and that it allows for unexpected solutions.

All \emph{evolutionary synthesis} models, on the other hand, have to assume a
stellar IMF and a SFH, i.e. the time evolution of the SFR for the galaxy. They
use stellar evolutionary tracks or isochrones that have to be complete in
terms of all relevant stellar evolutionary stages. 

Both methods need stellar spectral libraries that also have to be complete in
terms of stellar effective temperatures $\rm T_{eff}$, surface gravities $\rm\log
(g)$, and metallicities [Fe/H].

\galev models are available for a range of stellar IMFs, including
\cite{Salpeter55}, \cite{Kroupa01}, and \cite{Chabrier03}.  Other
choices of the IMF can easily be customized.

For the case of a simple stellar population (SSP), i.e. a star cluster, the
SFH is a $\delta$-function, meaning that all stars are formed in a single
timestep. \galev models for SSPs of different metallicities
\citep{Kurth+99,Schulz+02,AndersFritze03,LillyFritze06} were shown to well
reproduce observed colours and spectral indices of star clusters as function
of age and metallicity. They also show a pronounced non-linearity at
metallicities close to and above the solar value.  Important results include
the findings that any colour-to-age or index-to-age
calibration/transformations are only valid \emph{at one metallicity} and that
any colour-to-metallicity or index-to-metallicity calibration/transformations
are only valid \emph{at a given age} \citep[cf.][]{Schulz+02}, and that
extrapolating observational relations beyond their calibration range lead to
significantly misleading results.

In \cite{AndersFritze03} we
demonstrated the importance of nebular emission lines and continuum for young
stellar populations and showed that they can account for as much as $50 - 60
\%$ of the flux in broad-band filters, in particular at low metallicities.

For the description of undisturbed galaxies, SFHs have been determined for
normal average galaxies of types E, S0, Sa, Sb, Sc, Sd (cf. Sect.
\ref{subsec:calib:gal}), that, in combination with a Salpeter IMF extending from
a lower mass limit of ${\rm 0.1~M_{\odot}}$ (roughly the hydrogen burning limit)
to an upper mass limit around ${\rm 70 - 120~M_{\odot}}$, depending on the set
of isochrones (cf. Sect. \ref{subsec:input:isos}) selected, provide agreement
with average observed galaxy properties in terms of colours, spectra,
luminosities, abundances, and gas content. We stress that our galaxy types are
meant to denote \emph{spectral types} and we caution that the one-to-one
correspondence between spectral and morphological types observed in the Local
Universe might not hold to arbitrarily high redshifts.

While in terms of spectral evolution of the integrated light of galaxies (or
star clusters) our models are comparable to other evolutionary synthesis models,
they go beyond those in that they also allow to describe and analyze resolved
stellar populations in terms of colour-magnitude diagrams (CMDs) and in that
they self-consistently describe the chemical evolution of the ISM in galaxies
together with the spectral evolution of the stellar populations (for the latter
see Sect.  \ref{subsec:code:cc}), allowing to realistically account for the
coexistence of stellar subpopulations of different metallicities observed in
local galaxies. 

Most of the aforementioned capabilities are not entirely new. Models of
  the photometric evolution of galaxies date back to
  \cite{Tinsley67,Tinsley68,Tinsley72}, the first spectroscopic models appeared
  roughly a decade later \citep{Bruzual83,Guiderdoni+87}. Models of the chemical
  evolution of galaxies \cite[e.g.,][]{Truran+71,Tinsley72,Matteucci+93} were
  capable of taking the increasing enrichment of subsequent stellar populations
  into account when computing colours \citep[see also, e.g.,][]{Matteucci+87,
    Fritze+89}, spectra \citep[e.g.,][]{Bressan+94, FritzeGerhard94a,
    FiocRocca97, Pipino+04} and line indices \citep[e.g.,][]{Weiss+95,
    Bressan+96}. \cite{Silva+98} were the first to include a radiative
    transfer code into their model and could hence extend the wavelength
    coverage into the (far-)infrared. However, only very few models
  \citep[e.g.,][]{Prantzos+94, Portinari+98} exist that take the metallicity
  dependence of stellar yields into account and hence merit to be called
  \emph{chemically consistent}. Those, unlike \galev presented here, mainly
  focus on the metallicity distribution in the solar neighbourhood and only
  derive a detailed chemical evolution but no spectral or photometric
  evolution.

\subsection{Chemical evolution of galaxies}
\label{subsec:code:chemev}
Modeling the chemical evolution of galaxies starts from a gas cloud with given
initial (e.g.  primordial) abundances and given mass. A modified version of
Tinsley's equations \citep{Tinsley68}, including detailed stellar yields, is
solved to study the chemical enrichment history of galaxies of different
spectral types. This requires knowledge of stellar yields,
i.e. production rates of different elements and isotopes, including
contributions from SN Ia, as well as stellar lifetimes as function of stellar
mass \emph{and metallicity}, that can be taken from nucleosynthesis and stellar
evolution models, respectively.

Closed-box models can be compared to models with specified in- and outflow
rates and abundances. We follow the chemical evolution of a large number of
chemical elements H, He, ... Fe, fully accounting for the time delay between
SF and the return of material in stellar winds, PNe, and SNe.

\subsection{Chemically consistent GALEV models for galaxies}
\label{subsec:code:cc}
Combining the chemical evolution of ISM abundances and the spectral evolution
of the stellar population thus allows for what we call a chemically consistent
treatment of both the chemical evolution of the ISM and the spectral evolution
of the stellar population in galaxies: we use input physics (stellar
evolutionary tracks, model atmospheres, stellar lifetimes and yields) for a
large range of metallicities and consistently account for the increasing
initial abundances of successive stellar generations.

Broad stellar metallicity distributions have been reported for the Milky Way
disk \citep{RochaMaciel98}, bulge \citep{Sadler+96,Ramirez+00}, and halo
\citep{Ak+07}, as well as for the nearby elliptical galaxy NGC 5128
\citep{Harris+99,HarrisHarris00}.

Depending on the SFH of the respective galaxy type and eventually its infall
rate, stars of different ages within a galaxy will have different
metallicities and obey an age-metallicity relation determined by their
galaxy's SFH.

An important consequence of this coexistence of stars with different
ages and metallicities is that stars of different metallicities and
different ages dominate the light in different wavelength regions. It has
severe implications for metallicity indicators defined in different wavelength
regimes, which cannot be expected to trace one and the same stellar
metallicity. It also affects some widely used SF indicators and modifies, e.g.,
the calibrations for SFRs from H$\alpha$ or [OII] fluxes, as well as from FUV
luminosities \citep[cf.][]{BickerFritze05}.

Our \galev code can model the spectral and chemical evolution of galaxies
with arbitrary IMF and SFH %with a time resolution of 4 Myr
over cosmological timescales, from the very onset of SF all through a Hubble
time.  In combination with a cosmological model we can follow the redshift
evolution of galaxies from the early universe until today \citep{Bicker+04}.
It also allows to directly study the impact of evolutionary corrections as
well as of the chemically consistent treatment as compared to using solar
metallicity input physics only \citep[cf.][]{KotullaFritze09a}.

\subsection{Colour magnitude diagrams}
Despite simplifications the CMDs are valuable tools to study systematic effects
of SFH recovery from observations. \cite{FritzeLilly07} and \cite{LillyFritze08}
investigated the accuracy of recovering SFHs from CMDs as a function of the ages
of various subpopulations. In addition, they compared their results with the
accuracies of recovering SFHs from integrated spectra, multi-band photometry, or
Lick indices. These systematic studies are essential for more distant unresolved
stellar populations. Moreover, since synthetic CMDs can be calculated in any
desired filter combinations, they can be used to optimize observational
strategies with respect to the optimal filter combination, e.g.  to disentangle
ages and metallicities of young, intermediate-age or old stellar populations.

\section{Input physics}
\label{sec:input}

In the following sections we will review the input physics we use for our
\galev models.

\subsection{Stellar evolutionary tracks and/or isochrones}
\label{subsec:input:isos}
Data for stellar evolution can be taken either from isochrones or stellar
evolutionary tracks, both having their advantages and disadvantages. \galev
models currently use the most recent consistent set of theoretical isochrones
from the Padova group \citep[ff]{Bertelli+94} for five different metallicities
$\feh=(-1.7;-0.7;-0.4;0.0;+0.4)$ and include the TP-AGB phase, the importance
of which was shown in \cite{Schulz+02}. In order to be able to fully account
for emission lines we also include the Zero Age Main Sequence (ZAMS) into our
models. The isochrone for this ZAMS includes stars up to $\rm 120\,M_{\odot}$
and is created from the unevolved first data points of the stellar
evolutionary tracks. For more details we refer the reader to
\cite{BickerFritze05}.

\begin{figure}
\includegraphics[width=\columnwidth]{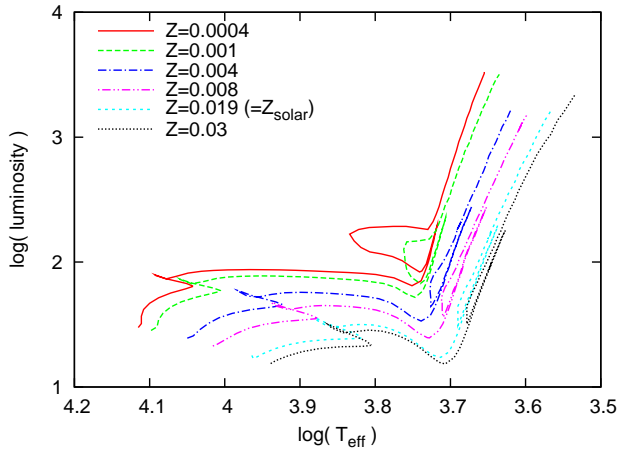}
\caption{Stelar evolutionary tracks of a $\rm 2\,\msun$ star for different
  metallicities ranging between $\rm Z=0.0004=1/50\,Z_{\odot}$ and $\rm
  Z=0.03=1.5\,Z_{\odot}$.}
\label{app:cc1}
\end{figure}

In Fig. \ref{app:cc1} we show stellar evolution tracks for a $\rm 2\,M_{\sun}$
star for 6 different metallicities ranging from $\rm Z=0.0004=1/50\,Z_{\odot}$
to $\rm Z=0.03=1.5\,Z_{\odot}$ \citep{Girardi+00}.  The general shape remains
unchanged with changing metallicity, but low-metallicity tracks are shifted
towards increasing luminosities and higher effective Temperatures, i.e. towards
the top left in the Hertzsprung-Russell diagram. This behaviour is the same for
stars of all masses, in the sense that with decreasing metallicity stars become
more luminous and their spectra shift towards higher effective temperatures. It
is therefore crucial to include those 
% non negligible 
effects to obtain a consistent picture of galaxy evolution.

\subsection{Library of stellar spectra}
\label{subsec:input:specs}
In principle, every library of stellar spectra -- observed or theoretical --
can be used, provided it is complete in terms of stellar ${\rm T_{eff}}$,
$\log(g)$, and [Fe/H]. \galev assigns spectra to stars in the isochrones
according to each star's metallicity, effective temperature $\rm T_{eff}$ and
surfce gravity $\log(g)$, normalising the spectra with the star's luminosity
(for details see Sect. \ref{subsec:program:isospec}). So far, \galev uses the
BaSeL library of model atmospheres from \cite{Lejeune+97,Lejeune+98},
originally based on the \cite{Kurucz+92} library. The wavelength coverage
spans the range from the XUV at $\rm\lambda \approx 90\,\AA$ to the FIR at
$\rm\lambda = 160\,\mu m$, with a spectral resolution of $\rm 20\,\AA$ in the
UV-optical and $\rm 50-100\,\AA$ in the NIR range. We remind the reader that
there are significant contributors other than starlight (e.g. PAHs, thermal
emission from cold dust) at wavelengths beyond the K-band, that are not
currently included in our models.  This wavelength range should hence be used
with caution.

\begin{figure}
\includegraphics[width=\columnwidth]{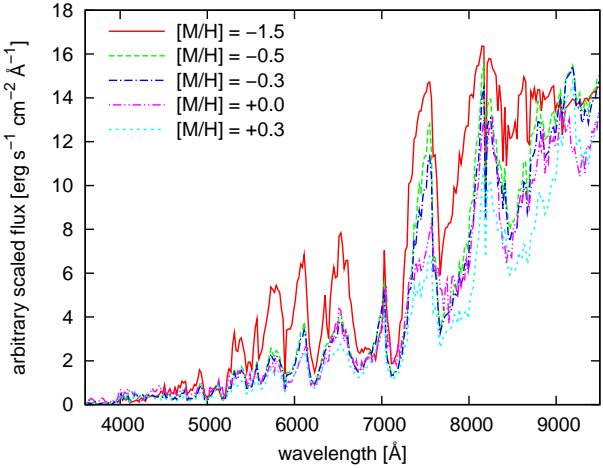}
\caption{Stellar spectra from the Lejeune library for a star with identical
  $\rm T_{eff}=3000K$ and $\log g=4.0$, but different metallicities from $\rm
  [M/H]=-1.5\ldots +0.3$. }
\label{app:cc2}
\end{figure}

Stellar spectra are heavily influenced by metallicity due to the increased
absorption line strength (e.g. for Fe-lines) and line-blanketing with
increasing abundances.  In Fig. \ref{app:cc2} we show the example of a cool
star ($\rm T_{eff}=3000K$, $\log g=4.0$) from the above mentioned BaSeL
library. The effects of metallicity are stronger for cool stellar atmospheres
where molecular absorption by VO, TiO, NH$_4$, H$_2$O etc. plays a larger role
\citep{Allard+00,Kucinskas+06}.

\subsection{Gaseous emission: lines and continuum}
\label{subsec:input:lines}
In addition to the stellar absorption spectra we also compute line and
continuum emission from gas ionized by hot massive stars. 

We do not take the ionizing photons from the stellar spectral library, but instead use
the tabulated values from \cite{Schaerer+97} for $\rm N_{LyC}$ as a function
of stellar effective temperature, radius, and metallicity. They yield much
better agreement with observations as well as with recent results from
expanding non-LTE, line-blanketed models atmosphere calculations by
\cite{Smith+02}.

We can then compute the total flux emitted by the gas per unit
wavelength as
\begin{equation}
F_{\lambda}=\frac{\gamma_{\lambda}(T)}{\alpha^{(2)}(T)}N_{LyC} =
\frac{c}{\lambda^2} \frac{\gamma_{\nu}(T)}{\alpha^{(2)}(T)}N_{LyC}
\end{equation}

with the speed of light $c$, electron temperature $\rm T = 10\,000\,K$, and
total recombination coefficient $\rm\alpha^{(2)}(10\,000\,K) = 2.575 \times
10^{-13}\,cm^{3}\,s^{-1}$ \citep{Aller84}. The gas continuum coefficients are
then computed for $\rm T = 10\,000\,K$ following \cite{ErcolanoStorey06} that
contains an algorithm to compute the bound-free radiation for hydrogen and
helium ($\rm \gamma_{HI}$, $\rm \gamma_{HeI}$, and $\rm \gamma_{HeII}$). The
HI two-photon emission coefficient $\rm \gamma_{H2p}$ is taken from
\cite{NussbaumerSchmutz84} and for the free-free emission $\rm \gamma_{b}$ we
use the formula from \cite{BrownMathews70} and compute the Gaunt factors
using the algorithm from \cite{Hummer88}. All $\gamma$-factors are summed to
form the final:
\begin{equation}
\rm \gamma_{\nu} = \gamma_{b} + \gamma_{H2p} + \gamma_{HI} +
\frac{n_{He^+}}{n_{H^+}}\gamma_{HeI} + \frac{n_{He^{++}}}{n_{H^+}} \gamma_{HeII}.
\end{equation}

For the densities of the helium $\rm He\,I$ and $\rm He\,II$ ions relative to
$\rm H\,I$ we use values typical for $\rm H\,II$ regions in galaxies
(Ercolano, priv. comm.), $0.0897$ and $1.667 \times 10^{-4}$ for $\rm He\,I$
and $\rm He\,II$, respectively. The final isochrone spectra are not very
sensitive to slight changes in these values.

The line strengths of the hydrogen lines are computed using atomic physics and
the emission rate of ionizing photons $\rm N_{LyC}$ from O- and early B-stars.
From the number of ionizing photons, $\rm N_{LyC}$, we compute the flux in the
H$\beta$ line using
\begin{equation}
\rm f(H\beta) = 4.757 \times 10^{-13} erg \times N_{LyC}.
\end{equation}
Assuming Case B recombination for a pure hydrogen cloud with $\rm
T_e=10\,000\,K$ \citep{Osterbrock+06}, we can then derive the line strengths of
all other hydrogen lines. Line strengths for heavier elements are computed using
metallicity-dependent line-ratios relative to H$\beta$. For metallicities $\rm
[Fe/H]\geq-0.4$ those are taken from the \cite{Stasinska+84} photoionization
models, adopting typical Galactic values of ${\rm T_e = 8100~K}$ and ${\rm
  n_e=1~cm^{-3}}$ for electron temperature and density. For lower metallicities,
we use observed line ratios from \cite{Izotov+94,Izotov+97} and
\cite{Izotov+98}.  These are supposed to include systematic changes in ${\rm
  T_e,~n_e}$ and the ionizing radiation field in lower metallicities
environments \cite[cf.][]{AndersFritze03}. Note that we do not include
  detailed radiation transfer calculations \citep[as, e.g., in][]{
    GarciaVargas+94,Ferland+98,Ercolano+03}, hence we caution the user to use
  line-ratios for detailed line diagnostics. We correct the gaseous emission
for small amounts of dust within the HII regions by reducing the ionizing flux
by 30\% if the gaseous metallicity is near solar, i.e. for $\feh\geq-0.4$. For
lower metallicities we use the full ionizing flux since those environments are
essentially dust-free \citep{Mezger78}. We do not account for yet unknown
  amounts of dust depletion of heavy elements within the HII regions. However,
  in the case of low-metallicity galaxies these effects are already included in
  the observed line-ratios.

Note that we do not account for internal self-absorption of Lyman
  continuum photons within the HII regions or the surrounding galaxy, since the
  fraction of flux leaking out of these regions is still a matter of ongoing
  debate \citep[see, e.g.,][]{Ferguson+96, Castellanos+02, FernandezSoto+03,
    Inoue+05, Siana+07, Wise+08}.

Depending on application, our models can include both continuum and line
emission, continuum emission only or no gas emission at all. In
\cite{Krueger+95} and \cite{AndersFritze03} we showed that in the case of Blue
Compact Dwarf galaxies and in young and metal-poor SSPs, gaseous emission can
contribute as much as 60\% to the flux in broad-band filters.  Emission lines
are the dominant contributors in the optical, whereas continuum emission
dominates in the NIR.

Combining the effects of higher luminosities and higher effective temperatures
with the effects of longer lifetimes of high-mass stars at low metallicities has
a profound impact on the spectrum of galaxies. All three factors lead to bluer
colours, higher overall luminosities and as a further aspect significantly
stronger gaseous emission, i.e. gaseous continuum and line emission.

\begin{figure}
\includegraphics[width=\columnwidth]{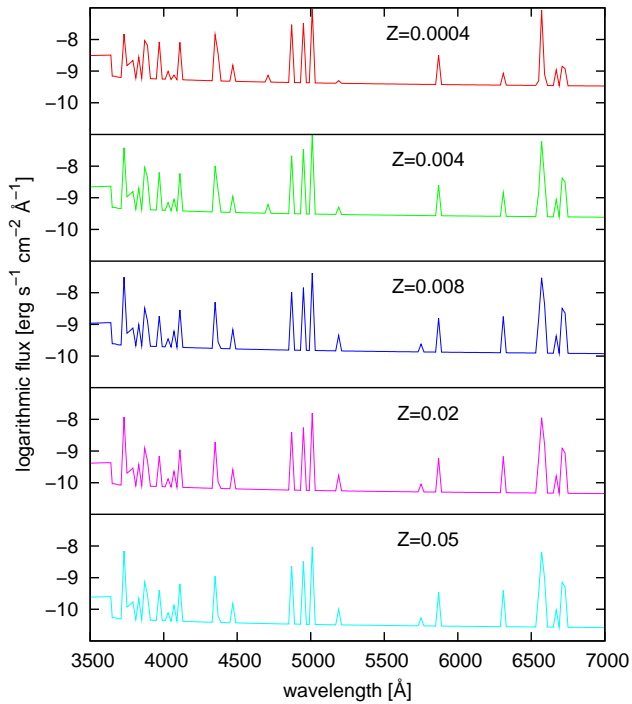}
\caption{Emission line spectrum of the $\rm 4\,Myr$ isochrone for 5 different
  metallicities ranging between $\rm Z=0.0004~([Fe/H]=-1.7)$ and $\rm
  Z=0.05~([Fe/H]=+0.4)$.}
\label{app:cc4}
\end{figure}

In Fig \ref{app:cc4} we show the gaseous emission spectra for isochrone spectra
at an age of $\rm 4\,Myr$ and for 5 metallicities from $1/50\,Z_{\odot}$ to $\rm
2.5\,Z_{\odot}$. The first notable aspect is the overall emission line strength
that is higher by a factor of about 10 comparing the two extreme cases. But
also the individual line ratios change significantly, i.e. the ratio between
H$\alpha$ (6563 \AA) and [N~II] (6583 \AA) ($\rm H\alpha/[NII]\approx 48$ at low
metallicity and $\approx7$ for solar metallicities). This extreme ratio is
directly affected by the lower nitrogen abundance at low metallicity. Other line
ratios, e.g. [O~II] (3727 \AA) to [O~III] (5007 \AA) also change due to the
more intense radiation field coming with the low metallicity environment.

\subsection{Lick stellar absorption indices}
\label{subsec:input:lick}
Since the resolution of the \cite{Lejeune+97,Lejeune+98} library is not
sufficient to calculate Lick absorption indices directly from the spectra,
\galev models use the empirical calibrations presented by \cite{Gorgas+93},
\cite{Worthey+94} and \cite{WortheyOttaviani97}. From those, \galev calculates
the fluxes in the Lick indices for every star, sums them up for the entire
stellar population at every timestep to yield integrated index fluxes and, in
combination with continuum fluxes from the integrated absorption line spectra,
the respective index equivalent widths \citep{Kurth+99,LillyFritze06}. \galev
at the present stage includes the following Lick indices: H$\delta_A$,
H$\gamma_A$, H$\delta_F$, H$\gamma_F$, CN$_1$, CN$_2$, Ca4227, G4300, Fe4383,
Ca4455, Fe4531, Fe4668, H$\beta$, Fe5015, Mg$_1$, Mg$_2$, Mgb, Fe5270, Fe5335,
Fe4506, Fe5709, Fe5782, NaD, TiO$_1$, and TiO$_2$ \cite[see][and references
therein for all index definitions]{Trager+98}.

\subsection{Stellar yields}
\label{subsec:input:yields}
To model the chemical enrichment histories of galaxies, \galev uses stellar
yields for a large number of individual elements (H, He, Li, Be, B, C, N, O, F,
Ne, Na, Mg, Al, Si, P, S, Cl, Ar, K, Ca, Sc, Ti,V, Cr, Mn, Fe, Co, Ni, Cu, Zn,
Ga, Ge) from \cite{Woosley+95} for massive stars and from \cite{vandenHoek+97}
for low-mass stars of various metallicities. Stellar lifetimes are taken from
the isochrones. There is a minor inconsistency in doing so, since
  \cite{Woosley+95} used models without mass-loss, while current isochrones in
  general account for mass-loss. However, since yields are only available for a
  very coarse metallicity grid this does not significantly affect the resulting
  chemical evolution. \galev also includes type Ia SN yields for the carbon
deflagration white dwarf binary scenario (W7) from \cite{Nomoto+97}. See
\cite{Lindner+99} for a detailed description of the chemically consistent
chemical evolution aspects of \galev.

\subsection{Ejection rates and remnant masses}
\label{subsec:input:remnant}
One of the central input parameters for \galev is the time-dependent ejection
rates necessary to compute the chemical evolution of galaxies. Those rates are
derived from the initial stellar masses $\rm M_{\star}$ and the remaining
remnant masses $\rm m_R$. For stars with masses $\rm M_{\star} \geq 30\,\msun$
the remnant is assumed to be a black hole of $\rm m_{BH}=8.0\,\msun$, with the
remaining mass being returned to the ISM; stars with initial masses of $\rm
30\,\msun \geq M_{\star} \geq 6.0\,\msun$ result in a neutron star of mass
$\rm m_{NS}$ given by \cite{Nomoto+88}:
\[
\rm m_{NS}[M_{\odot}] = 1.02 + 3.6363 \times 10^{-2} (M_{\star}/\msun - 8.0)
\msun;
\]
For the mass range of $\rm 6.0\,\msun \geq m_{\star} \geq
0.5\,\msun$, for which the stellar remnant is a white dwarf, we use a fit to
the data points of \cite{Weidemann00}:
\[
\rm m_{WD}[M_{\odot}] = 0.444 + 0.0838 (M_{\star}/\msun);
\]
This, combined with the extrapolation of the NS relation down to $\rm
6\,\msun$, provides a better matched connection between the two mass ranges.
while being compatible with the slightly steeper slope derived by
\cite{Kalirai+08}. However, as the remnant is only used to derive the mass
returned to the ISM during its lifetime and death, the exact transition point
from neutron star to white dwarf is of minor importance.

Stars with masses $\rm m_{\star} \la 0.5\msun$ have lifetimes in excess of a
Hubble time and negligible winds, hence do not return any material to the ISM.

\begin{figure}
\includegraphics[width=\columnwidth]{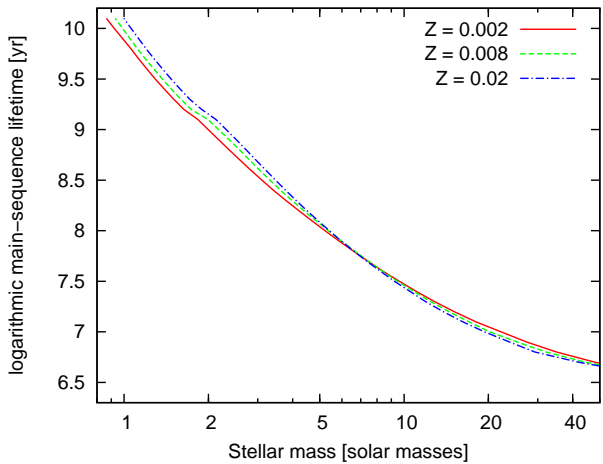}
\caption{Main-sequence lifetimes as function of initial stellar mass for three
different metallicities.}
\label{app:cc3}
\end{figure}

In Fig. \ref{app:cc3} we show the main-sequence lifetimes of stars as function
of initial mass for three different metallicities from $\rm
Z=0.002=1/10\,Z_{\odot}$ to $\rm Z=0.02= Z_{\odot}$, based on data from
\cite{Marigo+08} for $\rm M_{\star}<7\,\msun$ and \cite{Bertelli+94} for $\rm
M_{\star}>7\msun$.  Low-mass stars live longer at high metallicities, while
high-mass stars have longer lifetimes for lower metallicities; stars with
masses $\approx 3-4\,\msun$ have roughly the same lifetimes independent of
metallicity (see Fig. \ref{app:cc3}). The $\rm 20\msun$ star for example has a
$10\%$ increased lifetime at $Z=1/10\,Z_{\odot}$ as compared to $\rm
Z=\,Z_{\odot}$.

\subsection{Filter functions and magnitude systems}
\label{subsec:input:filter}
\galev includes a large number of filter functions to be convolved with the
model spectra in order to avoid uncertainties associated with transformations
between filter systems. It also provides the option to choose the desired
magnitude system Vegamag, ABmag, and STmag to be directly comparable to
observations and avoid transformations between different magnitude systems.
Magnitudes in the Vegamag systems are defined to have magnitude zero for an
A0V star; we use the Vega-spectrum from the \cite{Lejeune+97,Lejeune+98}
library combined with the flux calibration from \cite{BohlinGilliland04b} as
our standard star. AB magnitudes \citep{Oke74,BohlinGilliland04b} are derived
from the monochromatic flux $f_{\nu}$ such that $\rm m_{AB} =
-2.5\log(f_{\nu}) + 48.6$ if $f_{\nu}$ is measured in $\rm
erg\,s^{-1}\,cm^{-2}\,Hz^{-1}$; a colour of 0 in the AB magnitude means that
the object emits constant flux per unit \emph{frequency} interval; analogous
to the above, colours of 0 in the ST magnitude systems mean constant emitted
flux per unit \emph{wavelength} interval. The zeropoint has been chosen such
that a source with $\rm f_{\lambda}=3.63\times10^{-9}
erg\,s^{-1}\,cm^{-1}\,\AA^{-1}$ has $\rm m_{ST}=0\,mag$ in all filters.

Our current database contains filters from all major telescopes and
instruments, including all HST instruments (incl. WFC3), many ESO instruments,
and all common filter sets like Johnson/Cousins, Str\"omgren, Washington, and
SDSS.  However, note that due to the wide wavelength sampling of $\rm 20\,\AA$
of the current spectral library, narrow-band filters cannot be adequately
supported at this stage.

\subsection{Dust extinction}
To account for extinction and reddening due to interstellar dust, \galev also
implements the most commonly used empirical extinction laws from
\cite{Calzetti+00} and \cite{Cardelli+89}. The earlier was derived from actively
star-forming galaxies and describes a relatively gray extinction without the
$\rm2175\,\AA$ bump, characteristic for the latter extinction law.  For the
Cardelli law we assume a mean extinction parameter of $\rm R_V=A(V)/E(B-V)=3.1$,
characteristic for relatively quiescent galaxies.  Although models exists
  that offer a more detailed treatment of dust extinction and in some cases dust
  emission \citep[see, e.g.,][]{Silva+98, Popescu+00, Cunow01, Cunow04,
    Dopita+05, Moellenhoff+06, Piovan+06a, Piovan+06b}, the major drawback of
  these models is that they in general require assumptions on the geometry and
  spatial distribution of dust, gas and stars. However, the aim of \galev is to
  describe the spectrum of the average represantation of each galaxy type,
  making if difficult to compare these two approaches.

\subsection{Cosmological model}
\label{sect:cosmology}
\galev can also be coupled to a cosmological model to describe the
  evolution of galaxies as function of redshift. For this purpose we have to
  convert ages into redshifts and vice versa. In its current version we
  implemented a flat cosmology ($\rm \Omega_K = 0$). The choice of the local
  Hubble constant $\rm H_0$, the density parameters $\rm\Omega_{M}$ and
  $\rm\Omega_{\Lambda}$ (with the additional constraint $\rm
  \Omega_{M}+\Omega_{\Lambda}=1$) and the galaxy formation redshift ${\rm
    z_{form}}$ then completely determine the galaxy age as function of
  redshift. Because of the short time interval between e.g. ${\rm
    z=12~and~z=5}$, the exact value of ${\rm z_{form}}$ has very little impact,
  hence we choose an intermediate value of $\rm z_{form}=8$ as our default.

\subsection{Intergalactic attenuation}
\label{subsec:input:att}
To correctly describe spectra of higher redshift galaxies one has to include
the absorption shortwards of the Lyman-$\alpha$ line due to intervening
neutral hydrogen clouds. For that reason \galev implements the description for
the average attenuation effect as function of redshift following
\cite{Madau95} and covering the range $0\leq\rm z \leq 7$.

\section{Program structure}
\label{sec:program}
The actual modeling process with \galev can be divided into three steps:
\begin{enumerate}
\item In the first step \galev convolves the isochrones of each
  age-metallicity combination with the specified IMF, normalized to
  $1\,\msun$.  It then assigns a spectrum from the stellar library to each
  star on each isochrone and computes the integrated isochrone spectrum. Then
  the gaseous continuum emission and emission lines are added to the young
  isochrone spectra.  Using the yield tables and stellar remnant masses
  described in Sect. \ref{subsec:input:yields} with stellar lifetimes from the
  isochrones, \galev also derives the gas and metal ejection
  rates needed for the chemical evolution.
\item In the second step \galev computes the chemical and
  spectral evolution of the desired stellar population. For each timestep the
  current isochrone spectra are weighted with the SFH. The contributions from
  older SF episodes are obtained by integration over all past timesteps. The
  required interpolations in age and metallicity are described below.
  \galev thus calculates the time evolution of the integrated spectrum
  and its resulting line strength (i.e. Lick-) indices of a simple (SSP, star
  cluster) or composite (galaxy) stellar population, including the gaseous
  emission where appropriate. From the ejection rates and the available gas
  mass \galev calculates the new gaseous metallicity that will be used for
  stars born during the next timestep.
\item 
  In the last step \galev converts the integrated spectra into
  magnitudes in a large number of filters. Given a list of requested filters
  it convolves the spectra with the filter functions and applies zero-points
  to yield absolute magnitudes for each timestep. It can alternatively be
  combined with a cosmological model to yield apparent and absolute magnitudes
  as function of redshift. If requested it also accounts for dust extinction
  using observed extinction laws and, in the context of cosmological evolution,
  also for the attenuation by intergalactic neutral hydrogen.
\end{enumerate}
This process is also explained in a more vivid step-by-step explanation in
appendix \ref{app:pictorial}.

\subsection{Computation of isochrone spectra}
\label{subsec:program:isospec}
A crucial step is the assignment of stellar spectra from the library to the
points describing the isochrones. Since the available stellar parameters $\rm
T_{eff}$ and $\rm \log(g)$ in the library often do not match those required by
the isochrones, \galev has to interpolate between them in both
$\rm T_{eff}$ and $\rm \log(g)$. For a given combination of $\rm T_{eff}$ and
$\rm \log(g)$ this is done as follows:
\begin{enumerate}
\item Find up to four spectra bracketing the required values in both $\rm
  T_{eff}$ and $\rm \log(g)$.
\item Interpolate the spectra to the required value of $\rm T_{eff}$ in each
  pair of lower (upper) values of $\rm \log(g)$, yielding two new interpolated
  spectra with the correct $\rm T_{eff}$ and different $\rm \log(g)$. An
  important factor during this interpolation is the weighting of the spectra
  with each star's luminosity given by the isochrone. We choose to use the 
  integrated luminosity in the Johnson-V or Bessell-H-band, depending on the
  temperature of the star given by $\rm T_{eff}$. The original approach to
  normalize all stars in the V-band turned out to be insufficient for cool
  giants with only little flux in the optical. A ``cool giant'' in this
  context is defined by $\rm T_{eff} \leq 3500\,K$ and $\rm \log(g) \leq 3.5$.
\item The spectrum for the required value of $\rm \log(g)$ is then obtained by
  interpolation between the two spectra with the right $\rm T_{eff}$, again
  weighting with each star's respective luminosity.
\end{enumerate}
For the very hot stars ($\rm T_{eff} > 50\,000\,K$), the BaSeL stellar library
does not provide spectra. In those cases we extend the spectral library by
approximating the missing stellar spectra with black-body spectra of the
requested temperatures. The validity of this approximation for wavelengths
  longwards of $\rm\lambda\approx230\,\AA$ is supported by only minor
  differences between pure black-body and true spectra from both observations
  \citep{Gauba+01} and modelling \citep{Rauch03} of very hot central stars of
  planetary nebulae.

In a final step, all isochrone spectra are normalized to a distance of $\rm
10\,pc$, and are given in units of $\rm erg\ s^{-1}\ cm^{-2}\ \AA^{-1}$.

\subsection{Interpolation of isochrone-spectra and integration of galaxy
  spectra}
\label{subsec:program:interpol}
One important aspect of \galev is how to interpolate between the
isochrone spectra of different ages and metallicities, in other words how to map
the coarse grid of isochrones available onto the finer grid needed for galaxies.
\galev here interpolates logarithmically between ages and linearly
between metallicities expressed as $\log(Z)\sim\feh$. During their very early
stages the gaseous metallicities of galaxies are lower than the metallicity of
the lowest metallicity isochrone with $\feh=-1.7$.  For stars born at these
stages, we chose to use the lowest metallicity isochrone to avoid the
uncertainties in an extrapolation to lower metallicities. The same is done for
stars born from gas with a metallicity higher than the highest metallicity
isochrone. For these stars we use the highest metallicity isochrone available
with $\feh=+0.4$.

\section{Calibration of the GALEV models and comparison to observations}
\label{sec:calibration}
We stress that with the input physics as outlined above, a stellar IMF with
lower and upper mass limits chosen and total mass of the galaxy or star
cluster specified, \galev models provide the time evolution of
spectra, luminosities, and colours \emph{in absolute terms}. The same holds
true for the gas content and the chemical enrichment of galaxies. \emph{No a
  posteriori} gauging is applied.

The only exception is the cosmological context for galaxies: before
luminosities and magnitudes of redshifted galaxies are calculated, the B-band
model luminosities after a galaxy age corresponding to redshift $\rm z=0.0044$
(i.e. the redshift of the Virgo cluster) in the chosen cosmological model are
scaled to the average observed B-band luminosity of the respective galaxy type
in the Virgo cluster (cf. Sect.  \ref{subsec:calib:gal}).

In the following we present as an overview the comparison of \galev models for
star clusters and for normal galaxy types E, Sa $\ldots$ Sd with observations 
and refer to previous papers for more details.

\subsection{Star clusters}
\label{subsec:calib:sc}
The colours from U through K predicted by \galev models for SSPs or star clusters
at an age of $\sim 12 - 13$ Gyr are in very good agreement with the respective
observed colours for a large set of M31 and Milky Way globular clusters
\citep{Barmby+00a,Barmby+00b} at their respective metallicities
\cite[cf.][]{Schulz+02}. Small deviations in the U- and B-bands can be explained
by the existence of Blue Straggler stars (likely products of stellar mergers) in
the dense cores of GCs, which can make significant contributions at those short
wavelengths \citep{Xin+07,XinDeng05,Cenarro+08}, but are not included in our
models yet, as standard isochrones exist for non-interacting single stars
only. Further contaminants in the blue-to-UV region are Blue Horizontal
  Branch stars \citep{Schiavon+04,Dotter+07} and to a lesser degree even
  Cataclysmic Variables and low-mass X-ray binaries in the far-UV range
  \citep{Rich+93,Dieball+07}.

As shown in \cite{Kurth+99} and \cite{LillyFritze06}, Lick indices for
SSPs as calculated by \galev agree well with Galactic and M31 globular cluster
data as compiled by \citep{Harris96} in his online
catalog\footnote{http://physwww.physics.mcmaster.ca/{\textasciitilde}harris/mwgc.dat}.

\galev models for star clusters also show good agreement at an evolutionary age
of $\sim 12 - 13$ Gyr with the empirical calibrations for colours $(B-V)$ and
$(V-I)$ versus metallicity [Fe/H] for old Galactic and M31 globular clusters,
as e.g. given by \cite{Couture+90} and \cite{Barmby+00a} over the metallicity
range ${\rm -2.3 \leq [Fe/H] \leq -0.5}$ of these clusters. They also show,
however, significant deviations from a linear extrapolation of these empirical
relations towards higher metallicities, and they also show that the empirical
relations are only valid for the \emph{old} globular clusters for which they
have been derived. Models can, of course, be used to study the behaviour of
colour -- metallicity relations for any colour and as a function of time
% in their respective time evolution 
\cite[cf.][for details]{Schulz+02}.  

\begin{figure}
\includegraphics[width=\columnwidth]{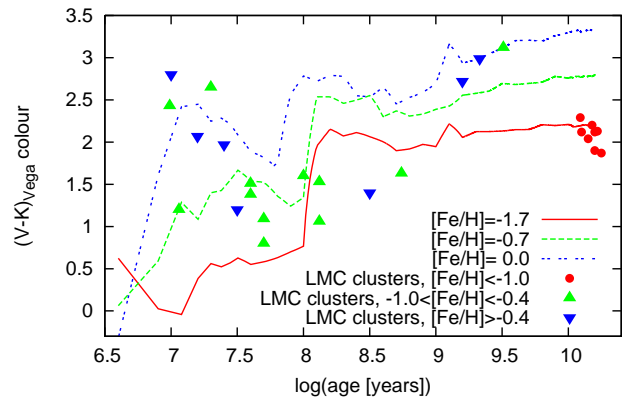}
\caption{Evolution of the V-K colour of our SSP models with metallicities
    of $\rm [Fe/H]=-1.7$ (red, solid line), $\rm [Fe/H]=-0.7$ (green, dashed),
    and $\rm [Fe/H]=0.0$ (blue, dash-dotted). We also plot the data (ages from
    \citet{Mackey+03}, colours from \citet{Persson+83}) for 35 clusters in the
    Large Magellanic Cloud for comparison.}
\label{fig:lmcclusters}
\end{figure}

In Figure \ref{fig:lmcclusters} we plot the evolution of three SSP models
  with different metallicities ranging from the lowest available value of $\rm
  [Fe/H]=-1.7$ to the solar value $\rm [Fe/H]=0.0$. For comparison we also show
  V-K colours from \cite{Persson+83} for star clusters in the Large Magellanic
  Cloud, colour-coded according to their metallicity as derived from CMDs
  \cite{Mackey+03}. Our models are able to not only reproduce the full range of
  observed colours, but also match the colour-evolution of each metallicity
  subpopulation if accounted for typical uncertainties of $0.4$ dex in the
  age-determination. For the old and metal-poor globular clusters our predicted
  V-K colours are in good agreement with observations.

As shown in \cite{LillyFritze06}, \galev models for SSPs also agree, after
$12-13$ Gyr of evolution with empirical calibrations of Lick indices
vs. [Fe/H]. However, they also show that \emph{all metal-sensitive indices are
  also age-dependent} and that the famous \emph{age-sensitive H$_{\beta}$ and
  H$_{\gamma}$ indices are also metal-dependent} to some extent \citep[see
  also][]{Thomas+03,Korn+05}. Hence, the empirical calibrations of Mg$_2$, Mgb,
etc. vs. [Fe/H] are valid only for \emph{old} (i.e. $>10$ Gyr) globular
clusters.

Empirical calibrations for colours or Lick indices vs. metallicity or age should
therefore not be used for star clusters or globular clusters for which it is not
a priori clear that their properties fall within the range of the calibrating
Galactic clusters. Instead a full set of colours and/or indices in comparison to
an extended grid of models covering the full parameter space, allows for
independent and accurate determinations of both ages and metallicities
(cf. Sect. \ref{sec:app:sc}).
 
\subsection{Galaxies}
\label{subsec:calib:gal}
By default, all our \galev models for galaxies use a Salpeter IMF
\citep{Salpeter55} with lower and upper mass limits of $0.1\,\msun$ and
$100\,\msun$, respectively.

\begin{figure}
\includegraphics[width=\columnwidth]{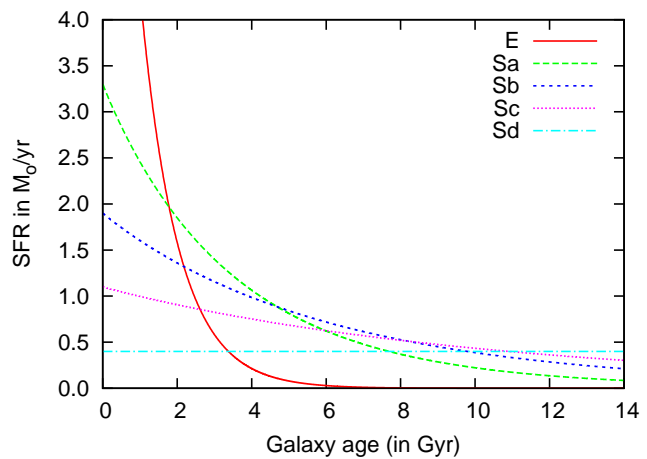}
\caption{Star formation histories for the five different galaxy types E and Sa
  through Sd. Absolute numbers are normalized to a total galaxy mass of
  $10^{10}\msun$.}
\label{fig:sfhs}
\end{figure}

The star formation histories, i.e. the time evolution of the star formation
rates, of the different spectral galaxy types E and Sa to Sd are \emph{the}
basic parameters of \galev and, in fact, of any kind of evolutionary synthesis
models. In Fig. \ref{fig:sfhs} we show SFHs for galaxies of different types
from E through Sd, assuming the same total (i.e. stars and gas) mass ${\rm
  M_{tot}}$ of $\rm 10^{10}\,\msun$ for all of them. These SFHs are in good
agreement with chemical and spectrophotometric findings from \cite{Sandage86}.

For the elliptical model we use an exponentially
declining SFR
\begin{equation}
\rm \Psi(t) = \frac{M_{tot}}{\alpha} \times \exp(-t/\tau)
\label{eq:E}
\end{equation}

with an e-folding time $\rm\tau=1\,Gyr$ and $\alpha=8.55\times10^{8}$ yr.

For spectral types S0 and Sa through Sc the SFR is tied to the evolving gas
content as given by
\begin{equation}
\rm \Psi(t) = \beta \times \frac{M_{gas}(t)}{10^9},
\label{eq:Sabc}
\end{equation}

with efficiency parameters $\beta$ decreasing from early to later spiral
types, ${\rm \beta=1.0\,yr^{-1}}$ for S0, ${\rm \beta=0.33\,yr^{-1}}$ for Sa, ${\rm \beta=0.19\,yr^{-1}}$ for Sb,
and ${\rm \beta=0.11\,yr^{-1}}$ for Sc galaxies.

Sd model galaxies are described by a constant SFR:

\begin{equation}
\rm \Psi(t) = \psi_0 \times \frac{M_{tot}}{10^{10}} = const,
\label{eq:Sd}
\end{equation}
where ${\rm \Psi_0=0.4~M_{\odot}\,yr^{-1}}$.

For \emph{closed-box models}, these parameters $\alpha$, $\beta$, $\psi_0$,
and $\tau$ are the only free parameters. We do \emph{not} require additional
parameters such as infall of gas or outflow in galactic winds to reproduce
observations. Note that all these SFH parameters are independent of galaxy
  mass. We therefore do not reproduce mass-metallicity or colour-magnitude
  relations for galaxies of identical spectral types.

These SFHs are very similar in \emph{all} evolutionary synthesis models
\citep*[cf. e.g.][]{BruzualCharlot03,FiocRocca97}. In detail, we adjust the SFH
for the chemically consistent \galev models as to match, after a Hubble time of
evolution (i.e. $\rm\approx13\,Gyr$ for our assumed concordance cosmology with
$\rm H_0=70\,km\,s^{-1}\,Mpc^{-1}$, $\rm\Omega_M=0.30$ and
$\rm\Omega_\Lambda=0.70$), the observed
\begin{itemize}
\item average integrated colours from UV through NIR, 
\item average gas fractions,
\item average metallicities,
\item average present-day SFRs,
\item average mass-to-light ratios, and 
\item template UV -- optical spectra
\end{itemize}
of the respective spectral types as detailed below. All these observational
constraints together very neatly define the average SFHs of undisturbed
galaxies E, Sa, Sb, Sc, Sd and tightly constrain the few parameters describing
them for a given IMF.

We stress that \galev and all other evolutionary synthesis models with their
respective SFHs are meant to describe \emph{spectral types} of galaxies. And we
caution that while in the Local Universe and for undisturbed galaxies a clear
one-to-one correspondence is observed between spectral and morphological
types, it is an open question how far back in time this correspondence might
hold.

\medbreak\textbf{Gas fraction:} The above described parameters have been tuned
to reproduce the typical gas fractions observed in local galaxies
\citep[e.g.][]{ReadTrentham05}. We use gas contents, defined as fractions of gas
relative to the total (gas+stars) mass, of ${\rm M_{gas}/M_{tot} = 0.0}$ for
E-models, 0.05 for Sa, 0.15 for Sb, 0.30 for Sc, and 0.55 for Sd models,
respectively. 

\medbreak\textbf{Colours:} With these parameters being fixed, we compare a wide
range of model-predicted colours from near-UV to near-IR to values from the
literature.  Our (B-V) colours, e.g. of 0.86 (E), 0.78 (Sa), 0.64 (Sb), 0.56
(Sc), and 0.43 (Sd) compare very well with the ranges found in the RC3 catalog
of \cite{deVaucouleurs+95}, listing mean colours of $\rm
(B-V)=0.89^{+0.17}_{-0.53}$ (E), $0.74^{+0.25}_{-0.35}$ (Sa),
$0.66^{+0.22}_{-0.47}$ (Sb), $0.51^{+0.22}_{-0.50}$ (Sc), and
$0.44^{+0.22}_{-0.18}$ (Sd).

\begin{figure}
\includegraphics[width=\columnwidth]{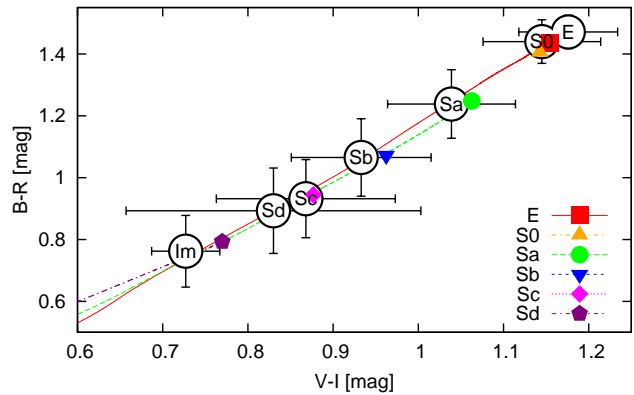}
\label{fig:galscolcol}
\caption{Evolution of galaxies of different types E through Sd in the
  (B-R)--(V-I) colour-colour-plane. Colours after 13 Gyr are plotted with
  large symbols. We also show colours for typical galaxies (large circles
  around galaxy type) taken from \citet{Buta+95} and \citet{ButaWilliams95}.}
\end{figure}

\begin{figure*}
\includegraphics[width=\columnwidth]{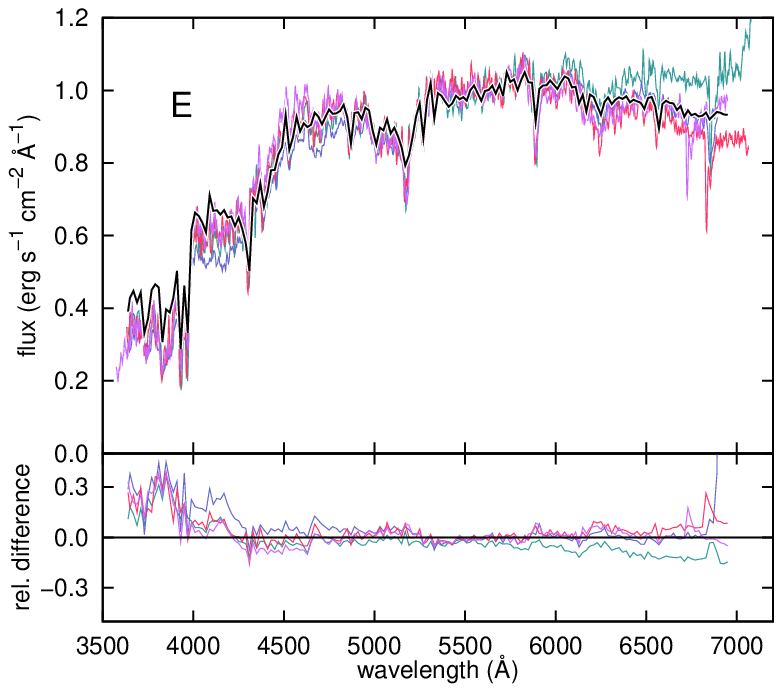}\hfill
\includegraphics[width=\columnwidth]{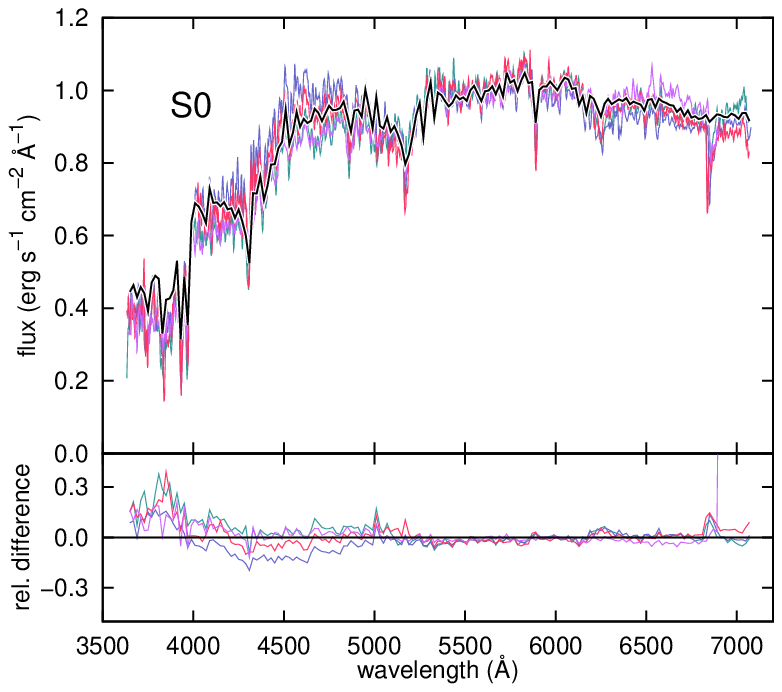}\\
\includegraphics[width=\columnwidth]{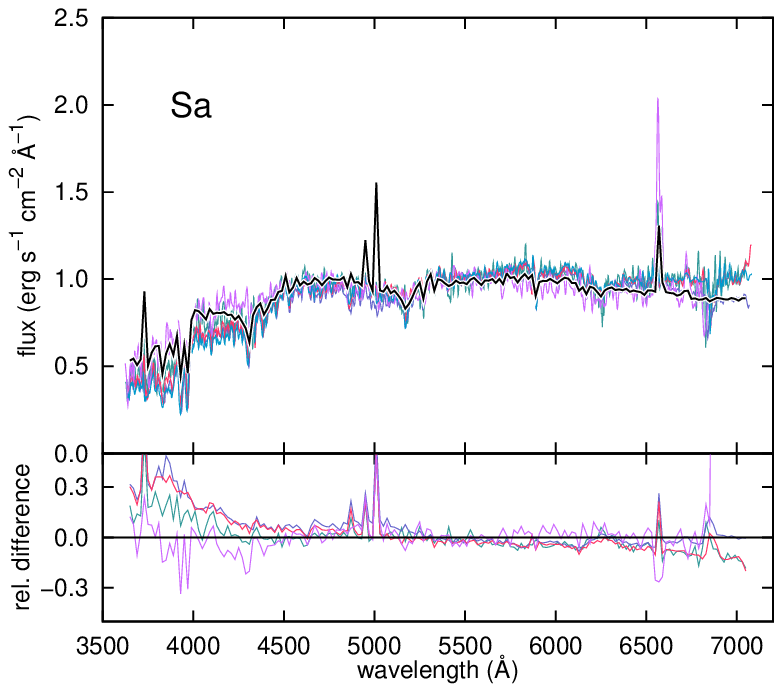}\hfill
\includegraphics[width=\columnwidth]{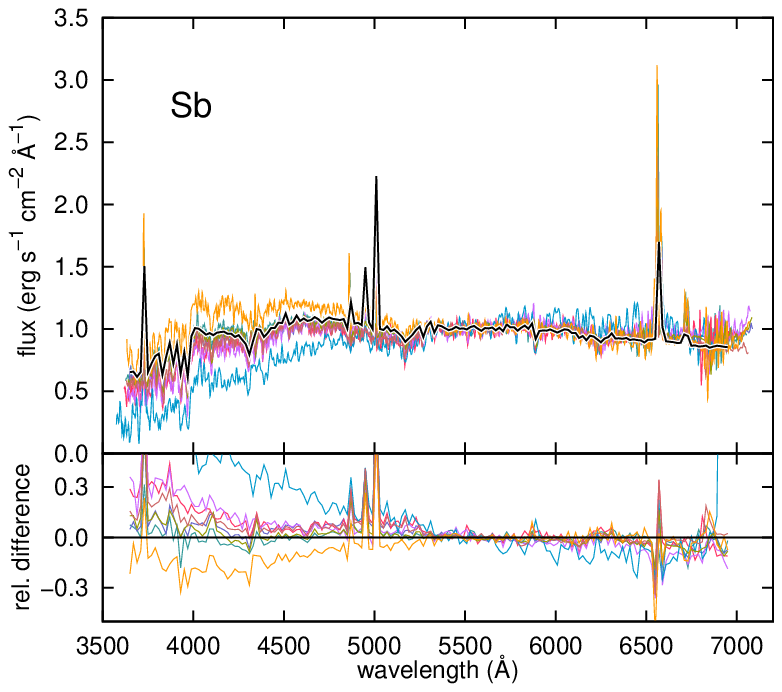}\\
\includegraphics[width=\columnwidth]{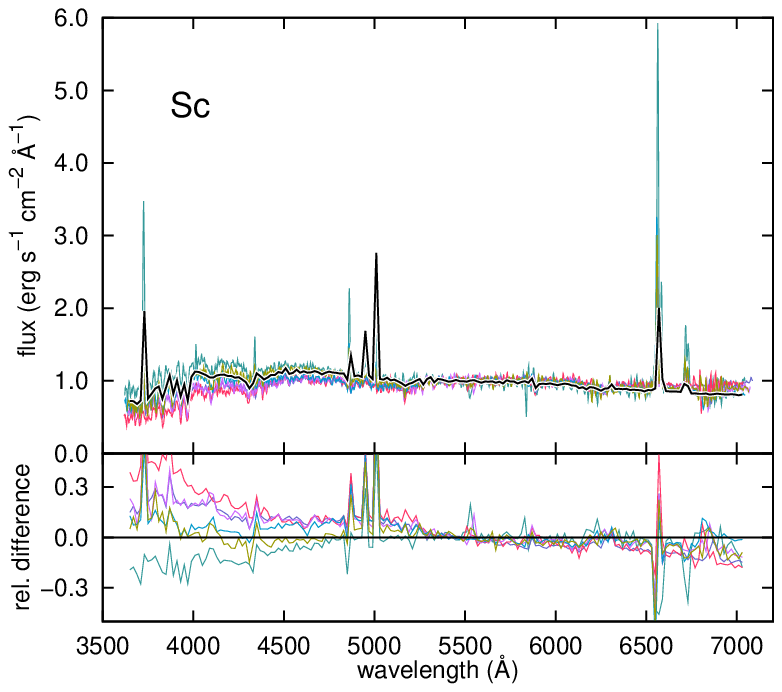}\hfill
\includegraphics[width=\columnwidth]{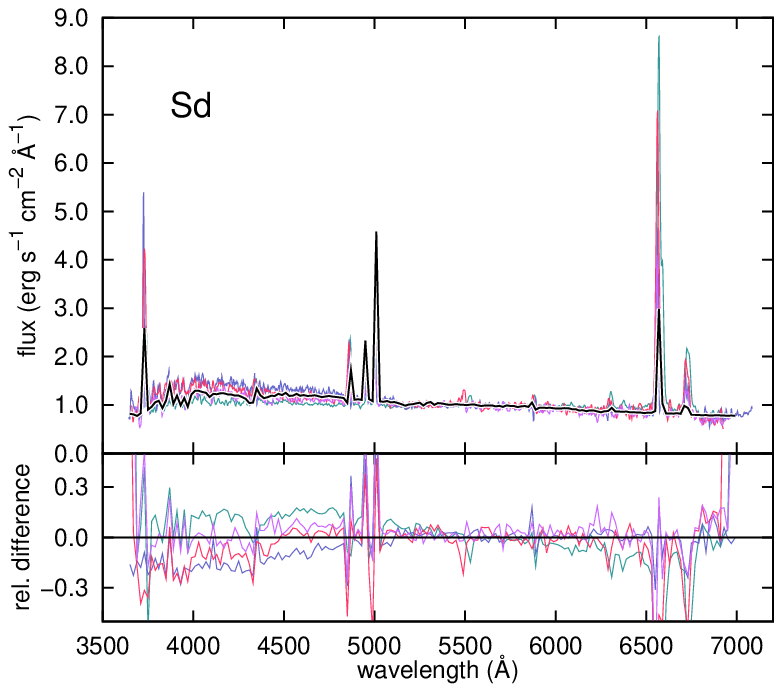}\\
\label{fig:specs}
\caption{Comparison of model spectra of different types (top row: E and S0,
  middle row Sa and Sb, bottom row Sc and Sd) with observed template spectra
  from \citet{Kennicutt92}. The black solid lines show the spectrum of
    \galev models at an age of 12.8 Gyr, the coloured lines show spectra of
    several template galaxies of each respective type. In the lower part of each
plot we also show relative differences between the \galev spectra and each
template spectrum.}
\end{figure*}

\medbreak\textbf{Spectra:} In Fig. \ref{fig:specs}, we compare our model spectra
to local templates for the galaxy types E, Sa, Sb, Sc, Sd from the catalog of
\cite{Kennicutt92}.  As can be seen from those plots, the observed spectra for
all spectral galaxy types from old passive ellipticals through the actively star
forming Sd galaxies are well reproduced. Although our spectra (here using
Lejeune's library) have a lower spectral resolution than the observed templates,
they nonetheless reproduce all features, like absorption and emission lines.
In the lower part of each plot we also show relative differences between
  model and template spectra. Deviations are generally smaller than differences
  between different galaxies of the same type, confirming the good
  agreement. Small differences for the emission lines originate in different
  spectral resolutions of templates and model spectra. Note that due to our
chemically consistent treatment \galev models can reproduce the Sd template
spectrum with our Sd model at an age of $\rm 13\,Gyr$. This is a notable
difference to all other evolutionary synthesis models that can only reproduce Sd
template spectra with younger models of ages $4-6$ Gyr
\citep[cf.][]{BruzualCharlot93}.

\medbreak \textbf{Metallicities:} As mentioned above \galev models calculate the
time evolution not only of individual element abundances but also of the global
ISM metallicity Z. After $\sim 13$ Gyr of evolution (for details on the
evolution as function of age or redshift see \citet{Bicker+04} or
\citet{KotullaFritze09a}), our models reach ISM abundances
of $\rm Z_{E}=\zsun$, $\rm Z_{Sa}=1.5\,\zsun$, $\rm Z_{Sb}=0.8\,\zsun$, $\rm
Z_{Sc}=0.5\,\zsun$, and $\rm Z_{Sd}=0.25\,\zsun$.  Those metallicities are in
good agreement with typical ISM abundances, as measured at $\rm 1\,R_{eff}$, the
effective or characteristic radius, of spiral galaxies of various types
\citep{Oey+93, Ferguson+98a, vanZee+98}. Note that our models aim to represent
the \emph{typical $L^\star$ or $M^{\star}$ galaxy of each spectral
  type}. Galaxies of different masses within the same spectral type are known to
have slightly different metallicities, as described by the mass-metallicity or
luminosity-metallicity relations \citep{Skillman+89, Tremonti+04, Kewley+08}. As
stated before, these relations are currently not accounted for by \galev standard
models. However, the user is free to choose parametrizations of the SFHs
  as a function of galaxy mass \citep[as, e.g., in][]{Bressan+94} that reproduce
  these relations.

\medbreak\textbf{Star formation rates:} \galev models predict the following
present-day SFRs $\Psi_p$ in [${\rm M_{\odot}\,yr^{-1}}$] for the different
spectral galaxy types: $\Psi_p \approx 0$ for E, $\Psi_p = 0.65$ for Sa, $\Psi_p
= 0.4$ for Sb, $\Psi_p = 0.47$ for Sc, and $\Psi_p = 0.13$ for Sd galaxies.
These numbers are in agreement with the average star formation rates of $0.22$,
$0.32\pm0.2$, $0.33\pm0.14$, and $0.09\pm0.03\,{\rm M_{\odot}\,yr^{-1}}$ for Sa
to Sd galaxies from \cite{Kennicutt+83b}. We caution that the smooth SFHs in
\galev and other evolutionary synthesis models certainly are simplifications;
real galaxies will in general experience fluctuations in their SFRs around these
mean values. For the Milky Way, \cite{Rocha+00} showed on the basis of
individual stellar ages and abundances that the mean global SFH indeed fell
nicely between those of our Sb and Sc models, with fluctuations of a factor of 2
on timescales of $\rm 100\,Myr$. Such fluctuations around the mean, however, do
not significantly affect the long-term evolution.

\medbreak\textbf{Mass-to-light ratios:} Using the stellar mass assembled after
$\sim 13$ Gyr and the absolute V-band luminosity, we compute the V-band
mass-to-light ratios. The results are: $\rm M/L_V(E,Sa,Sb,Sc,Sd)=11.8, 8.2, 5.9,
4.5, 3.0$, respectively. These include a factor that we call \emph{fraction of
  visible mass} (FVM) and that accounts for the fraction of mass locked up in
sub-stellar objects like brown dwarfs, planets, etc. This factor is set to
FVM=0.5 in \galev models, implying that only half the star formation rate forms
luminous stars, while the remaining half is locked up in non-luminous objects
\citep[cf.][]{Bahcall+92}. Only with this FVM=0.5 can agreement with observed
average M/L ratios \citep{FaberGallagher79,BelldeJong01,ReadTrentham05} be
achieved together with agreement in colours, spectra, and chemical
abundances. Note that FVM does not include remnants of stellar evolution like
white dwarfs, neutron stars or black holes; those are considered separately as
part of stellar evolution.

Introducing the FVM parameter does not only affect the mass-to-light
  ratios but also has an impact on the chemical evolution with in turn impacts
  on most other derived parameters. A Salpeter IMF with mass-limits of 0.1 and
  100 $\rm M_{\sun}$ would lock up 60\% of the mass in long-lived stars with
  $\rm M\le 1\,M_{\sun}$, with the remaining $40\%$ progressing the chemical
  enrichment of the galaxy. However, by applying a FVM=0.5 we half this fraction
  of stars with $\rm M\ge 1\,M_{\sun}$ so that now only $20\%$ of stars return
  their ejecta at the end of their lifes, with the majority of mass being locked
  up. This concept to slow down the chemical enrichment is a common feature of
  many codes, but comes with different names, e.g. as free parameter $(1-\xi)$
  in \citet{Portinari+98}.

\medbreak\textbf{Luminosity normalization} With the SFHs as given above and a
Salpeter IMF, \galev models calculate the time evolution of spectra and
luminosities in absolute terms as a function of the total mass, initially all in
gas. Colours, of course, are independent of galaxy mass as \galev models
assume the SFH of each galaxy type to be independent of galaxy
  mass (see Fig. \ref{fig:sfhs}). Chemical ISM abundances are also
determined by this SFH \emph{and} by the absolute level of SFRs in relation to
the mass in the initial gas reservoir.  Spectral fluxes and luminosities on the
other hand are mass-dependent. If models of undisturbed galaxies are to be
compared to observations of individual galaxies, it is important to calibrate
their absolute luminosities to the observed ones, thereby also calibrating their
(gaseous + stellar) masses and their SFRs.

Before we can compare models to distant galaxies, we need to calibrate their
apparent magnitudes in the Johnson B-band against average observed magnitudes
of the respective spectral galaxy types in the Local Universe. The average
observed apparent luminosities of the different galaxy types in the Virgo
cluster are given by \cite{Sandage+85e,Sandage+85d} to be $\rm m_B(E)=10.20$,
$\rm m_B(S0)=12.60$, $\rm m_B(Sa)=11.95$, $\rm m_B(Sb)=12.90$, $\rm
m_B(Sc)=12.83$, and $\rm m_B(Sd)=13.95$. The correction factors to match the
computed and observed luminosities are then also applied to all mass-dependent
model parameters, i.e.  to stellar and gaseous masses, SFRs, etc.

\subsection{Disturbed galaxies: Starbursts \& SF truncation}
\label{sect:starbursts}
In addition to the undisturbed galaxies above with their smooth and
monotonically decreasing SFRs, \galev can also describe the evolution of
galaxies encountering fast changes in their SFR, e.g. a starburst with
significantly enhanced star formation rate, or a truncation or strangulation
of the star formation rate on some shorter or longer timescale, respectively.
This allows us to study not only the spectral but also the chemical evolution
of galaxies that experience starbursts or a quenching of their normal SFR, as
they can occur e.g. during galaxy-galaxy interactions or mergers or in the
course of the various transformation processes discussed for field galaxies
that fall into galaxy clusters.  We recall that \galev models only deal with
effects on the SFRs of galaxies and should be viewed as complementary to 
dynamical models describing the morphological transformations eventually
related to these processes. Both starbursts and SF truncation/strangulation
require additional free parameters. The first of these is the time of onset
$t_b$, i.e. the age at which a previously undisturbed galaxy begins to feel
changes in its SFR. Starbursts are described by a prompt increase in SFR
followed by an exponential decline on some given timescale $\tau_b$, that
typically is of the order of a few hundred Myr for normal-size galaxies and of
order $10^5 - 10^6$ yr for dwarf galaxies. 

Across the literature, a number of different definitions for the strength of a
burst are in use.  The strength of a burst is either described by the amount of
stars formed during the burst as compared to the stellar mass at the beginning
or at the end of the burst or by the amount of all gas available at $t_b$ that
is transformed into stars during the burst, equivalent to the star formation
efficiency (SFE) of the burst. We prefer the latter definition and define the
burst strength via the star formation efficiency of the burst:

\begin{equation}
\label{eqn:burststrength}
\rm SFE(burst) = 
\frac{stellar~mass~formed~during~burst}{
%   -----------------------------------------------------
      gas~mass~available~at~onset~of~burst}
\end{equation}

This definition limits burst strengths to be within the range $0\ldots1$ and
gives, at the same time, a measure of the global SF efficiency during the burst.
Please note that since the burst strength depends on the amount of gas available
at the onset of the burst, bursts of the same strength but occuring at different
times and/or in galaxies of different types do not necessarily form the same
amount of \emph{stars}. E.g. a strong burst in an old and gas-poor Sa galaxy
will form a much smaller mass of new stars than a weak burst early in the life
of a gas-rich Sd galaxy. All burst strengths can only be determined at the end
of the bursts. Before the end of the burst only lower limits can be estimated.

A truncation or strangulation of the SFR, e.g. as a galaxy falls into a galaxy
cluster and has its HI stripped by ram pressure, is described either by an
abrupt termination of SF or an exponential decline of the SFR on a timescale
$\tau_b$. After a burst or after SF truncation/strangulation, \galev models can
have SFRs going either exponentially to zero or to some constant low level.

\section{The web interface}
We present the new web interface that enables everyone to run her or his model
of choice in a fast, easy to learn and easy to use, and comfortable way over the
internet. The interface can be found at
\begin{center}
\textbf{http://www.galev.org}
\end{center}
Using a web interface enables us to impose some checks previous to program
execution dealing with the most serious and frequently occuring problems.
For the user this has the important advantage of always having the latest
version at hand, including the most up-to-date input physics like isochrone sets
and stellar libraries.

\subsection{How to use the web-interface}
The web-interface consists of four steps: Two for the input of the
model parameters and the requested output, one for parameter checking and at
last the actual running of the program on the web-server.

\subsubsection{Principal model parameters}
On the first page the user has to decide which type of model (s)he wants to
run. Those defining parameters are

\begin{enumerate}
\item \textbf{Galaxy type}: choose between the most common types E, S0, and Sa
  to Sd for undisturbed galaxies, for which the parameters described in Sect.
  \ref{subsec:calib:gal} will be used. In addition we offer \emph{free} types,
  defining the shape of the SFH as described by eqs. \ref{eq:E}--\ref{eq:Sd}.
  We also offer instantaneous burst (SSP) models, as well as completely
  definable SFHs. For the latter mode the user has to specify a file with SFRs
  as function of time in the next step.
\item \textbf{Burst}: we currently offer three possibilities: \emph{No burst},
  i.e. an undisturbed galaxy, a \emph{Burst} with given strength, duration,
  and exponentially declining SFR, beginning at some time $t_b$ and \emph{SF
    Truncation} with the SFR declining exponentially to zero on some specified
  timescale, again from time $t_b$ on.
\item Number of filters for which magnitudes are to be computed.
\end{enumerate}
If the user chooses to run a standard model for an undisturbed galaxy, i.e. an
elliptical or spiral galaxy, then \galev in the following uses the parameters
described in Sect. \ref{subsec:calib:gal}. For all other cases, i.e. either
user-defined galaxy types or models with burst/truncation, the parameters will
be entered in the following step.

\subsubsection{Model configuration}
Depending on the choices from the previous page the user might not have all
possible parameters to choose from; only the ones required for the specified
model type will be shown. Those are:

\begin{enumerate}
\item We offer a set of \textbf{Initial Mass Functions (IMF)}, namely from
  \cite{Salpeter55} and \cite{Kroupa01}, both using the Lejeune
  \citep{Lejeune+97,Lejeune+98} stellar atmosphere library. Further IMF shapes
  \citep[e.g.][]{Chabrier03} and different sets of stellar libraries are in
  preparation. In a future release we plan to also allow the user to choose
  customized IMFs; this feature, however, will be limited to SSP models (i.e.
  for star clusters).

\item \textbf{Gaseous emission} can be switched off or on, and the user can
  choose to include only continuum emission or both continuum and line
  emission. In fixed metallicity models, the gaseous emission is evaluated for
  the respective metallicity, in the chemically consistent case, it is
  calculated appropriately as described in Sect. \ref{subsec:input:lines}.

\item The \textbf{metallicity}: choose between \emph{chemically consistent}
  models, i.e. those including all enrichment effects described above, or a
  model with metallicity fixed to one of the following values
  $\feh=-1.7,-0.7,-0.4,0.0,+0.4$.

\item The \textbf{galaxy type} is shown as a reminder, but cannot be changed
  here any more. To change the type the user has to go back to the first step.

\item \textbf{Galaxy mass} is another free parameter, mainly used for
  normalization. We do not include a mass-metallicity relation of any kind, so
  this parameter has no impact on the resulting colours. Note that when \galev is
  combined with a cosmological model to e.g. obtain apparent magnitudes, the
  model galaxy mass can optionally gets rescaled to match, after a Hubble time,
  the average B-band luminosity of local galaxies of the respective type in
  Virgo.

\item \textbf{SFH parameters}: if (and only if) one of the free models has
  been chosen, then the variables $\alpha$ (normalization for the SFR of an
  elliptical model), $\tau$ (e-folding timescale of an elliptical model),
  $\beta$ (factor of proportion for spiral models), and $\psi_0$ (constant SFR
  of an Sd model) can be freely chosen by the user. The meaning of those
  parameters is explained in more detail in Sect. \ref{subsec:calib:gal}. If
  in the previous step the user requested the SFH to be read from a file this
  will be specified here. We caution the reader to carefully study the
  web-output from \galev, since depending on the chosen parameters there might
  be problems, e.g. a SFR exceeding the amount of gas that is available.

\item If the user wants to compute a galaxy featuring a \textbf{burst} or SF
  \textbf{truncation}, then (and only then) he/she has to specify the time of
  the onset of the burst, expressed in years after galaxy formation and the
  e-folding timescale for the decline of the SFR during the burst.  While those
  two parameters are common to both truncation and burst, only in the latter
  case one also needs to supply a burst strength.  Following our definition
  of the burst-strength in Sect.  \ref{sect:starbursts} (eq.
  \ref{eqn:burststrength}), this specifies the fraction of gas available
  before the burst that is to be converted into stars during the burst. If one
  instead wants to specify the SFR at the very onset of the burst, $\rm
  SFR(t=t_b)$, those two numbers can be converted into each other by $b =
  SFR(t=t_b) \times \tau_b / M_{gas}(t=t_b)$. 

\item \galev currently supports two different \textbf{extinction} laws that can
  be applied to the spectra, the \cite{Calzetti+94} law for starburst
  galaxies, and the \cite{Cardelli+89} Milky Way extinction law, using the
  standard value of $R_V=3.1$. To compute colours for different extinction
  values, the user can specify a maximum E(B-V) value and a step-size. The
  output (see below) will then include multiple files, one for each extinction
  value in the requested range. 

\item \textbf{Cosmological parameters} need to be specified to model the
  evolution of galaxies with redshift. The user can choose Hubble constant
  $\rm H_0$, $\rm \Omega_{M}$, and $\rm \Omega_{\Lambda}$, while $\rm
  \Omega_{K}$ is fixed to 0. To convert ages into redshifts the user also has
  to select a formation redshift $\rm z_{form}$, so that the age of the galaxy
  is given by $\rm age(z) = t_H(z) - t_H(z_{form})$ with $\rm t_H$ being the
  Hubble time at redshifts z and ${\rm z_{form}}$. Note that the modeling
  process is terminated at a galaxy age of $\rm 16\,Gyr$. A choice of
  cosmological parameters leading to a galaxy age of more than $\rm 14\,Gyr$
  (e.g. too low a Hubble constant) hence results in an error-message, so that
  no colours will be computed for this case.

\item Although \galev offers a full range of different output options
  (see below), not all the numbers will be actually needed for any specific
  application. The user can therefore use the section \textbf{Output
    parameter} to restrict the output, leading to faster execution and smaller
  downloads.

\item To compute magnitudes from the spectra the user can choose from a large
  list of \textbf{filter functions} and also specify the magnitude system on
  which magnitudes are to be based (Vega, AB or ST magnitudes) 
  For each filter a different magnitude system can be chosen. This eliminates
  the uncertainties involved in any \textit{a posteriori} transformations
  between different systems.
\end{enumerate}

We also offer an extensive online help giving examples for different
parameters and how they affect the resulting SFH.

\subsubsection{Output}
In the following we will describe which outputs are available and also how
they are computed.

\begin{enumerate}
\item The \textbf{integrated spectra for each timestep} are the most
  important output, since from those all magnitudes and redshifted spectra are
  derived.
\item Convolving the spectra with the appropriate filter functions and
  applying the appropriate zeropoints gives us the \textbf{magnitudes} of
  the model as a function of time. 
\item \galev also delivers the most useful diagnostic data for each timestep,
  such as current \textbf{stellar and gaseous mass}, \textbf{star formation
    rates}, \textbf{ISM metallicity}, and ionizing flux $\rm N_{LyC}$.
\item After applying a cosmological model to convert galaxy age into redshift
  (assuming a formation redshift) we can compute the redshifted galaxy
  spectrum, that includes all evolutionary effects. We also apply the
  intergalactic attenuation from \cite{Madau95} to account for absorption of
  flux shortwards of Ly$\alpha$ by intervening neutral hydrogen clouds.
\item Convolving those redshifted spectra with filter functions ${\rm
    FF(\lambda)}$ and adding the bolometric distance modulus gives us apparent
  magnitudes with and/or without attenuation, that then can be used e.g. for
  comparison with observed SEDs to derive photometric redshifts or ages and
  metallicities.
\item From the data described above we can derive cosmological corrections due
  to the shifting of the filter functions to shorter restframe wavelengths
  (k-corrections) and evolutionary corrections (e-corrections). Those
  corrections are computed as follows:
  \begin{equation}
    \rm k(z) = -2.5 \times \log \frac{\int_0^\infty f(t=t_0,z,\lambda) \times
      FF(\lambda) d\lambda}{\int_0^\infty f(t=t_0,0,\lambda) \times FF(\lambda)
      d\lambda}
  \end{equation}
  \begin{equation}
    \rm e(z) = -2.5 \times \log \frac{\int_0^\infty f(t=t(z),z,\lambda) \times
      FF(\lambda) d\lambda}{\int_0^\infty f(t=t_0,z,\lambda) \times FF(\lambda)
      d\lambda}
  \end{equation}
  where $f(t,z,\lambda)$ is the flux at wavelength $\lambda$ of a galaxy of
  age $t$ at redshift $z$, $t_0$ the current age of the universe (which
  depends on the specified cosmological parameters), and ${\rm FF(\lambda)}$ a
  filter function.
\end{enumerate}

\subsubsection{Parameter checking and execution}
As a third step in the modeling process we perform a quick check to ensure that
all required parameters are given and correspond to valid combinations. The page
also displays all given parameters to allow the user to check the input, and
eventually perform corrections by going back to the previous page. However, the
tests performed are just basic validity tests and do not ensure that the
computed model makes physical sense.

In the fourth and last step, after the user has made sure that all input is
correct, we create all files necessary for the actual modelling. The execution
of the \galev program takes several minutes to run. We urge all users to
carefully read through the given output to check if everything ran smoothly.

Directly after \galev has created all spectra, it computes the magnitude
evolution as a function of time or redshift in all filters requested, again
taking a few minutes to complete.

For compatibility reasons all output is given as human-readable ascii-files,
with values aligned in columns that are separated by spaces, so that the files
can be easily analysed and/or plotted. The resulting files are then
automatically combined and compressed into one archive (.tar.gz) file, that
can be downloaded. Each archive also contains a small ReadMe file listing the
content of each file. The meaning of individual columns are specified at the
beginning of each file.

\subsection{Upcoming features of the web-interface}
In its current version the web-interface supports the most frequently used
\galev features like computing spectra and colours. Further features, like the
computation of Lick indices or colour magnitude diagrams are currently in the
process of being adapted and implemented into the webpage and will be
accessible online in the near future.

\section{Future prospects}
In the near future (Weilbacher et al. 2009, in prep.) we will offer
further stellar libraries, e.g. \cite{Munari+05} and \cite{Coelho+05},
to be able to compute high-resolution spectra for comparison with modern
spectroscopic surveys.
Those libraries have higher spectral resolution and shorter wavelength coverage
but are not colour-corrected in the way \cite{Lejeune+97,Lejeune+98} did for
the Kurucz spectra to reproduce the observed colours of stars from the UV
through the NIR over the full range of effective temperatures.

One drawback of current galaxy evolutionary synthesis models is that they do
not include a self-consistent treatment of dust absorption and reemission. One
factor contributing to this difficulty are geometric effects and dependencies
of dust masses and properties on gas content, chemical abundances and
eventually even radiation field. 
Early attempts to consistently include dust absorption as a function of gas
content and [Fe/H] in collaboration with D. Calzetti were encouraging
\citep[cf.][]{Moeller+01b} and showed that every undisturbed galaxy goes
through a phase of maximum extinction of $\rm E(B-V) \sim 0.4$. The redshift
of this maximum $E(B-V)$ phase is determined by the interplay between
decreasing gas content and increasing metallicity. The predicted values at
high-redshift $z\approx 3$ (equivalent to a galaxy age of $\approx\,$2 Gyr)
agree well with observations from \cite{Steidel+99}, \cite{Shapley+01} and
\cite{Colbert+06}.

Another aspect is the coupling of \galev to a dynamical model 
(SPH + N-body + SF + feedback) to cope with increasingly available data from
Integral Field Units allowing spatially resolved spectroscopy. Our early
attempt to couple \galev models for single stellar generations with a
cosmological structure formation simulation by M. Steinmetz was encouraging
\citep[cf.][]{Contardo+98} and showed that this approach is feasible. The
simulated HST images of a galaxy at different redshifts showed fairly good
agreement with observations but ultimately failed to reproduce the correct
local disk sizes and parameters. A better and more detailed description of
feedback seems to be required. This can only be obtained from an extensive
comparison between model results and resolved galaxy observations. The key
issue is to have a correct criterion for SF and a correct description of
feedback on the relevant scales and over the full range of SF activity -- from
the lowest levels in the farthest outskirts of galaxies to the highly clumped
and clustered SF in the strongest starbursts.

\section{Applications}
\galev models have a wide range of applications from star clusters and resolved
stellar populations of nearby galaxies through integrated properties of
galaxies up to the highest redshifts. A fair number of them have been explored
so far, many of them hand in hand with refinements or special features added
to the models.  Here we briefly recall a few of them to illustrate the various
features of \galev.

\subsection{Star clusters}\label{sec:app:sc}
The simplest stellar systems to study with \galev are star clusters, so-called
simple stellar populations (SSPs) where all stars are formed essentially within
one timestep and with the same chemical abundances. \galev models describe the
time evolution of SSPs with different metallicities, including the gaseous
emission during early evolutionary stages and as appropriate for their
respective metallicity. They can also incorporate extinction within the
clusters' parent galaxy on the basis of empirical extinction laws from
\cite{Calzetti+94} or \cite{Cardelli+89}.  

Using our AnalySED tool \citep{Anders+04a,Anders+04b}, we can compare
observations in widely spaced broad-/medium-band filters to a grid of \galev
models and derive physical cluster parameters such as ages, masses, metallicity,
and extinction between the cluster and the observer
If one of the parameters can be externally constrained (e.g. dust-free
environment, or metallicity previously determined from spectroscopy)
observations in at least 3 bands are required, otherwise at least 4 bands are
needed. As a large number of star clusters can usually be covered by a single
set of observations, this is a very efficient way to study statistically
significant cluster samples.

As shown both in studies based on artificial star clusters and on star clusters
with ages and metallicities derived independently from CMDs, accuracies in age
determination of ${\rm \Delta age/age \leq 0.3}$ and in metallicity
determinations of ${\rm \Delta [Fe/H]/[Fe/H] \leq 0.2}$ are achievable,
preferably if both a short-wavelength band (U or B) and a NIR-band are included
\citep{Anders+04b,deGrijs+03b,deGrijsAnders06}. The U-band is crucial for
accurate age-dating, and a NIR-band (H or K) for accurate abundances.  We
successfully applied these models in \cite{Anders+04a} to the interpretation of
the young star cluster systems in the starburst dwarf galaxy NGC~1569, in
\cite{Anders+07} to the analysis of the star clusters in the interacting
Antennae galaxies (NGC 4038/39), and in \cite{Kotulla+08a} to the derivation of
ages and metallicities of the globular clusters in the Virgo S0 galaxy NGC 4570.

\subsection{Colour-magnitude diagrams}
One special feature of \galev is its ability to compute colour-magnitude
diagrams (CMDs) in any desired passband combination. This is possible not only
for SSPs or instantaneous bursts, but also for composite stellar populations
with complex SFHs. This has successfully been used to identify the best
possible passband combination to disentangle age and metallicity effects in
star clusters in various age ranges \citep[cf.][]{Fritze+06}.

In \cite{FritzeLilly07} we compared the SFH obtained from the CMD with those
obtained from the integrated spectrum, from Lick index measurements and from
multi-band photometry in their respective accuracies and limitations. The basic
result was that none of the methods allows to look back beyond a recent burst or
some recent phase of enhanced SFR, and that all methods face very similar
accuracy limitations at look-back times beyond $\sim 1$ Gyr. Only within the
most recent Gyr, CMD analysis achieves the most detailed SFHs
\citep[cf.][]{LillyFritze05a,LillyFritze05b}. The importance of this kind of
comparative investigation lies in the fact that CMD analyses can only be done
for the resolved stellar populations within the nearest Local Group
galaxies. All attempts to explore the SFHs of more distant galaxies have to rely
on integrated spectra and, for the most distant ones, on integrated photometry
only.

\subsection{Undisturbed galaxies}
In \cite{BickerFritze05} we used \galev models to study the effects of the
chemical evolutionary state of galaxies on their star formation rate indicators
(H$\alpha$, [OII], NUV and FUV luminosities) and found that \emph{all of them} significantly depend on metallicity, with
errors in the worst cases of up to factors of a few, confirming previous
  observational evidence \citep[e.g.,][]{Jansen+01,Hopkins+03,Kewley+04}.

In \cite{Schulz+03} study the time and redshift evolution of bulge-to-disk light
ratios in different wavelength bands by assuming a short timescale for SF for
the bulge component ${\rm \Psi_{bulge} \sim \exp (-t/1\,Gyr)}$ and a constant
SFR for the disk component $\rm \Psi_{disk} \sim constant$. The integrated
spectral and photometric evolution of different spiral galaxy types was then
obtained by adding up the bulge and disk components in mass ratios so as to
give, after a Hubble time of evolution, the observed average B-band
bulge-to-disk light ratios for the respective spiral types Sa through Sd. This
study showed a significant wavelength dependence of the bulge-to-disk light
ratios in agreement with observations by \cite{Eskridge+02}. This has
implications for galaxy classification in different redshift intervals. It also
opens a new possibility to explore bulge formation scenarios and bulge formation
redshifts by comparing bulge-to-disk light ratios measured in different bands.

\subsubsection{Starbursts in Blue Compact Dwarf Galaxies}
Starbursts were first investigated with \galev models in the context of
Blue Compact Dwarf Galaxies (BCDGs) in a series of papers by H. Kr\"uger
\citep{Krueger+91,Krueger+92,Krueger+93,KruegerFritze94,Krueger+95}. These
authors investigated a sample of BCDGs with optical and NIR photometry in order
to derive their burst strengths and the age of their underlying stellar
population. The main results we want to recall here are that even very weak
ongoing bursts can completely dominate the light in the optical, in particular
at the low metallicities $Z_{\odot}/50 \ldots Z_{\odot}/5$
typical for BCDGs. An underlying old galaxy component can only be detected in
the NIR and was found for every BCDG of our sample. Very accurate age-dating was
possible for those BCDGs which showed a $\rm 4600\,\AA$ bump caused by WR-stars
in their spectra. Burst strengths were found to be of the order of a few percent
only, when defined in terms of stellar mass increase.  They were also shown to
systematically decrease with increasing galaxy mass, where the latter included
the important mass contributions of HI.  This result is in agreement with
expectations on the basis of the stochastic self-propagating SF scenario put
forward by \cite{GerolaSeiden78} and \cite{SeidenGerola79}.

\subsection{Interacting galaxies and mergers}
In \cite{FritzeGerhard94a,FritzeGerhard94b} we studied a grid of starburst
models with bursts of various strengths occurring in Sa, $\dots$, Sd spirals at
different ages in their spectral, photometric and chemical evolution and then
analyzed the starburst in the gas-rich massive spiral-spiral merger NGC 7252. We
found this burst to have started about $\rm 600 - 900\,Myr$ ago and to have been
stronger by $1-2$ orders of magnitude than those in BCDGs. The bulk of
information available for this galaxy even allowed us to estimate the SF
efficiency on the basis of a comparison of the stellar mass formed during the
burst.

Our conservative estimate for the overall SF efficiency (see equation
\ref{eqn:burststrength}) during this interaction-induced starburst indicated a
very high value ${\rm SFE \geq 0.35}$ \citep{FritzeGerhard94a, FritzeGerhard94b,
  FritzeBurkert95}, again about two orders of magnitude higher than any SFE
measured for molecular clouds in the Milky Way or the Magellanic Clouds, and
high enough to allow for the formation of a new generation of globular clusters
\citep[cf.][]{Brown+95,Li+04}. \galev models indicated that NGC 7252 at present
still features a low-level ongoing SFR of $\rm\sim 3\,M_{\odot}\,yr^{-1}$ in its
centre, powered for $\sim 50\,\%$ by gas set free at present from dying burst
stars and for $\sim 50\,\%$ by HI falling back onto the main body from the tidal
tails. Both of these gas delivery rates will decrease over the next
$\rm1-3\,Gyr$.  Depending on whether the SFR will cease completely or continue
at some very low level, NGC 7252 will spectrally evolve into an elliptical or S0
galaxy over the next $\rm 1-3\,Gyr$. Already at present, NGC 7252 features an
r$^{1/4}-$ light profile across a radial range of $\sim 14$ kpc, if azimuthally
averaged \citep{Schweizer82}

Independent confirmation for the unexpectedly high value found for the SFE in
NGC 7252 came from the detection of a rich population of massive compact star
clusters with ages in agreement with the global starburst age which, in turn, is
in agreement with dynamical merger ages from N-body $+$ SPH simulations
\citep[e.g.][]{HibbardMihos95}. Spectroscopy of the brightest clusters confirmed
their metallicities to be between $Z_{\odot}/2$ and $Z_{\odot}$, as expected if
they formed out of the gas pre-enriched in Sc-type spirals -- with some evidence
for a moderate amount of self-enrichment during the burst on the basis of their
slightly enhanced ${\rm [\alpha/Fe]}$ ratios.  In \cite{FritzeBurkert95} we
estimated that the number of clusters with masses in the range of Galactic
globular cluster (GC) masses that formed in the burst and survived until the
present is of the same order of magnitude as the number of GCs present in two
average-luminosity Sc-type spirals before the merger.  Hence, this spiral-spiral
merger will, after the fading of the post-starburst and after the fading and
dissolution of the tidal features, evolve into an elliptical or S0 galaxy with a
normal GC specific frequency.  GCs of age $0.5 \ldots 1$ Gyr have already
survived the most critical phase in their lives, the infant mortality and early
mass loss stages, and stand fair chances to survive for many more Gyr
\citep[cf.][]{Lamers+05, BastianGoodwin06, ParmentierFritze09}.

In \cite{deGrijs+03a} we used \galev SSP models to analyze the
luminosity-weighted ages pixel-by-pixel on ACS images of the interacting
galaxies Tadpole and Mice and studied their star cluster populations. A
surprising result was that about 35 \% by mass of all recent SF went into the
formation of star clusters in both galaxies, not only across the main bodies
of both galaxy systems but all along their very extended tidal tails.

In \cite{TemporinFritze06} we applied \galev models to investigate the SF and
starburst histories of galaxies in a very compact group of galaxies on the
basis of multi-band photometry and spectra and in \cite{Wehner+06} we studied
the SF activity and its history in the extended tidal debris surrounding the
starburst galaxy NGC 3310.

\subsubsection{Tidal Dwarf Galaxies}
Not only star clusters can form in the low-density environments of tidal tails
but sometimes even star-forming objects with masses in the range of dwarf
galaxies: so-called Tidal Dwarf Galaxies (TDGs), or better TDG candidates. In
\cite{Weilbacher+01} and \cite{Weilbacher+02,Weilbacher+03a,Weilbacher+03b} we
analyzed the first reasonably sized sample of TDGs and found that they all
contain a stellar population inherited from the spiral disk out of which the
tidal tail has been torn, together with a significant young stellar population
that must have been formed \emph{in situ} within the tidal tail after it had
been ejected. A characteristic feature of TDGs is that they do not follow the
luminosity-metallicity relation of dwarf galaxies but all have similar
metallicities characteristic of the HII region abundances in spiral disks. 

Again it turned out that optical observations alone are not sufficient to
disentangle the mass contributions of the inherited versus the starburst
components. Even a 90 \% mass fraction in the inherited component can be
entirely hidden in the optical by an ongoing burst that only makes up for 10 \%
of the mass. Only in optical-NIR colours can the inherited component be detected
that is not entirely old but contains the mix of stellar ages present in the
disk before the tidal tail was thrown out.

\subsection{Galaxy transformation in groups and clusters}
A variety of scenarios are discussed in the literature to explain the
transformation of the spiral-rich field galaxy population into the
S0-/dSph-/dE-rich galaxy population observed in rich galaxy clusters at low
redshift. The following processses have been proposed: 
High-speed disruptive galaxy-galaxy interactions called harassment, interactions
between galaxies and the dense hot Intracluster Medium (ICM), and enhanced
merging within infalling groups. All of these scenarios are observed to be at
work in a number of individual cases. Their relative importance, their
timescales, transition stages, and end products, however, are not known yet. All
of these transformation processes both affect the morphological appearance of
galaxies and -- via their SF histories -- their spectral properties. How the
timescales for morphological transformation and spectral transformation relate
to each other in the various scenarios and environments is not clear to
date. Removal of gaseous halos, outer, and inner HI disks leads to SF
strangulation on long timescales or to SF truncation on shorter
ones. Destabilization of disks through encounters or shocks as well as mergers
within infalling groups may lead to starbursts. We explored aspects related to
the spectral transformation of galaxies through all these scenarios and
investigated which scenarios in which type of progenitor galaxy and at which
evolutionary stage can lead to the observed luminosity and colour ranges of S0
galaxies by implementing SF strangulation/truncation on different timescales
with and without preceding starbursts into \galev models in \cite{Bicker+02}. In
\cite{Falkenberg+09a,Falkenberg+09b} we extended the models to
also include the evolution of the D4000 and H$\delta$ Lick indices into the
\galev models for galaxy transformation and investigated under which conditions
the so-called E$+$A-, or k$+$a-, and the H$\delta$-strong galaxies
\citep[cf.][]{Poggianti+04,Dressler+04} are formed, what is the lifetime of this
respective phase, what is the colour and luminosity of the galaxy in this
transition stage and what is the end-product.

\subsection{High redshift galaxies and photometric redshifts}
If coupled to a cosmological model, \galev can be used to study the evolution of
galaxies from the very onset of SF in the early universe until today.
Accounting for the significantly sub-solar metallicities observed at high
redshifts -- in particular when dealing with intrinsically faint galaxies that
dominate in deep field surveys -- allows us to determine more accurate
photometric redshifts, as compared to those obtained with solar-metallicity
models only \citep{KotullaFritze09a} or observed templates (Kotulla 2009, {\it in
  preparation}). 

In \cite{FritzeBicker06} we examined starbursts and their respective
post-starburst stages across a wide range of redshifts with the surprising
result that dust-free models in their long post-burst phases after strong
starbursts at high redshifts can get the colours and luminosities of Extremely
Red Objects (EROs) as e.g. observed in the K20-survey \citep[cf.][]{Daddi+02,
  Cimatti+02}.

\begin{figure}
\includegraphics[width=\columnwidth]{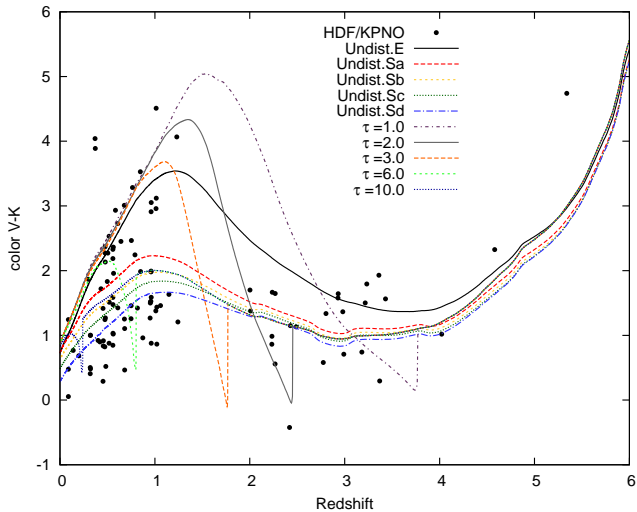}
\caption{V-K colour as function of redshift for five undisturbed spectral
  types E and Sa through Sd. Extremely blue colours are found during active
  starbursts, while extremely red colours can be reached during the
  postburst phases of early starbursts, here shown for galaxy ages of 1, 2, 3,
  5, and 10 Gyr.}
\label{fig:hdfvk}
\end{figure}

Only when we also include -- in addition to our set of undisturbed models E, Sa,
$\ldots$, Sd -- the very blue starburst phases and also their \emph{extremely
  red postburst phases} can we reproduce the full range of colours observed
e.g. in the Hubble Deep Fields (Kotulla \& Fritze 2009, in preparation). In Fig.
\ref{fig:hdfvk} we show the F606W-Ks (approx.  V-K) colour evolution for a set
of undisturbed models E, and Sa through Sd, and for models with major starbursts
(burst strengths chosen to consume 70\% of the in each case available gas)
occurring at different ages in a previously undisturbed Sb galaxy. A comparison
to the photometric galaxy catalog of \cite{FernandezSoto+99} for the HDF shows
that \galev models can describe the full colour range, even without any
dust. Note that dust certainly plays a role in ongoing starbursts but not any
more during postburst stages.  Apparently, as observed in the local Universe,
galaxies use up and destroy part of their dust in starbursts, while the rest is
blown away by the end of the burst.

\subsection{Redshift evolution of ISM abundances: spiral models vs. DLAs}
The SFHs of our closed-box model galaxies were constrained by requiring
agreement after 12$-$13 Gyr of evolution not only with observed
spectrophotometric properties of the various galaxy types but also with their
gas content and ISM abundances (cf. Sect. \ref{subsec:calib:gal}). While this
ensures agreement at the present stage it needs not necessarily imply agreement
over the entire evolutionary path. Some subtle interplay between SFR and infall
rate, e.g., could lead to the present-day agreement after ages of
disagreement. Infall rates are very hard to constrain from a comparison with
spectrophotometric properties alone. They can, however, be constrained
by comparing chemical abundances. It is therefore of prime interest to compare
the redshift evolution of ISM abundances to observations of high-redshift
galaxies. 

Transforming the time evolution of ISM abundances into a redshift evolution only
requires the one-to-one transformation between galaxy age and redshift that is
given by any set of cosmological parameters (see. Sect. \ref{sect:cosmology})

Damped Lyman-$\alpha$ absorbers (DLAs) are a particular class of absorption
systems, usually observed in the line-of-sight of quasars, with damping wings
on the Lyman-$\alpha$ line due to the high column densities of neutral
hydrogen of ${\rm N(HI) \geq 10^{21}\,cm^{-3}}$ in the absorbers. The damped
Lyman-$\alpha$ line is always accompanied by a number of low-ionisation heavy
element lines, many of which are the dominant ionisation states in those
absorbers. DLAs are most easily observed in the redshift range ${\rm z \sim
  2}$ to ${\rm z > 4}$ and have been proposed to be potential progenitors of
present-day spiral disks: Their HI column densities are similar to
those in local gas-rich disks and their gas content, derived from rotational
velocities of the order $100 \ldots 200 km\,s^{-1}$, is similar to
gaseous plus stellar masses in local spirals. In \cite{Lindner+99} we compared
the redshift evolution of our spiral galaxy models, calculated in a chemically
consistent way with stellar yields for a large number of elements and 5
different stellar metallicities to the first reasonably sized set of Keck
HIRES abundances for DLAs from the literature. We found surprisingly good
agreement for all the different abundances over the entire observed redshift
range with our closed-box models.

Studying the impact of various amounts of infall showed that it was not possible
to accommodate more than a moderate amount of infall, increasing the total mass
from redshift 2 until the present by not more than a factor $\sim 2$, without
losing agreement with either the spectral or the chemical properties or both, no
matter how we tuned the SFH. We hence concluded that from a chemical evolution
point of view DLAs might well be the progenitors of present-day spiral galaxies
of all types Sa $\ldots$ Sd and that those are reasonably well described by
closed-box or moderate-infall models. This implies that DLAs have already almost
all the mass of present-day spirals -- albeit almost completely in the form of
HI gas. \galev models indicated low stellar masses and low luminosities for DLA
galaxies, in agreement with the large number of non-detections/upper
limits. Mass estimates from rotation velocities derived from detailed modeling
of asymmetries detected in some DLA profiles confirmed our mass predictions
\citep[cf.][]{Wolfe+05}. \galev models also predicted a change in the DLA galaxy
population from high to low redshift and showed that the DLA phenomenon can be
understood as a normal transition stage in the life of every spiral. During
their enrichment process to higher metallicities they convert their gas
reservoir into stars, therefore get increasingly gas-poor
so that above a certain metallicity they drop out of DLA samples due to too
low a gas content \citep[cf.][for more information]{Lindner+99}. This is in
agreement with observations that show the lowest redshift DLA galaxies to be
low-luminosity late-type or irregular galaxies.

\section{Summary}
This paper presents the \galev evolutionary synthesis models for star clusters,
undisturbed galaxies and galaxies with starbursts or/and star formation
truncation now available on the web at 
\begin{center}
\textbf{http://www.galev.org}
\end{center} We describe the input physics currently used, that will
continuously be updated.

For a number of different stellar IMFs, the spectral evolution of star
clusters of metallicities in the range ${\rm -1.7 \leq [Fe/H] \leq +0.4}$ can
be calculated, and a large number of filter systems are available for the
photometric evolution as well as the full set of Lick absorption indices.

\galev features a unique combination of characteristics that
  allow for what we call a \emph{chemically consistent modelling} of the
  chemical evolution of the ISM together with the spectral evolution of the
  stellar component.

This means that the initial abundances of every stellar generation are accounted
for by using input physics (stellar evolutionary tracks, stellar model
atmospheres, gaseous line and continuum emission, stellar lifetimes, yields and
remnant masses) appropriate for the increasing initial abundances present at the
formation time of successive stellar generations. This chemically consistent
modeling accounts for the observed broad stellar metallicity distributions in
local galaxies as well as for the increasing importance of subsolar abundances
in local late-type and low luminosity galaxies and in high redshift galaxies.

Galaxy models can be calculated either in the chemically consistent way or for
some fixed metallicity upon request. Models give spectra, emission line
strengths and Lick absorption features, photometric quantities for a large
number of filter systems, and chemical abundances, gaseous and stellar masses,
star formation rates, etc. in their time evolution for normal galaxies,
galaxies with starbursts or/and star formation truncation as specified by the
user or for user customized star formation histories.

If a cosmological model is selected, all quantities are also provided in their
redshift evolution, fully accounting for evolutionary and cosmological
corrections and including the attenuation by intergalactic neutral hydrogen.

We present the models, the input physics they use, their calibrations, the web
interface and some examples of selected applications for illustration and we
discuss current limitations and future prospects.

\section*{Acknowledgments}
It is a great pleasure to thank all the former members of the \galev group in
G\"ottingen for their contributions. We thank all our collaborators for
inspiring suggestions -- and J. S. Gallagher, R. de Grijs, and P.-A. Duc in
particular. We also thank E. Brinks for a thorough read of the manuscript and
especially his idea to add the pictorial appendix. We thank B. Ercolano for help
with the gas continuum emission and our anonymous referee for her/his comments
and suggestions that helped us to clarify parts of this paper.

P.M.W. recieved financial support through the D3Dnet project from the German
Verbundforschung of BMBF (grant 05AV5BAA). We gratefully acknowledge continuous
support from the Deutsche Forschungsgemeinschaft.

\appendix

\section{Working in pictures}
\label{app:pictorial}
In the following we will present the workings of \galev in a more vivid way by
guiding the reader through a series of steps. We give examples for the input
physics \galev uses and show how this is used throughout the process of
modelling a galaxy.

\subsection{General steps}
In the first step we present how one gets from isochrones, a stellar IMF,
spectral library, and atomic physics to an integrated isochrone spectrum.  For
this example we choose to work with isochrones from the Padova group, a
Salpeter-IMF from $0.1\msun$ to $100\msun$ and stellar spectra from the Lejeune
library, all using solar metallicity.

\subsubsection*{Step 1: Choose set of isochrones}
\begin{figure}
\includegraphics[width=\columnwidth]{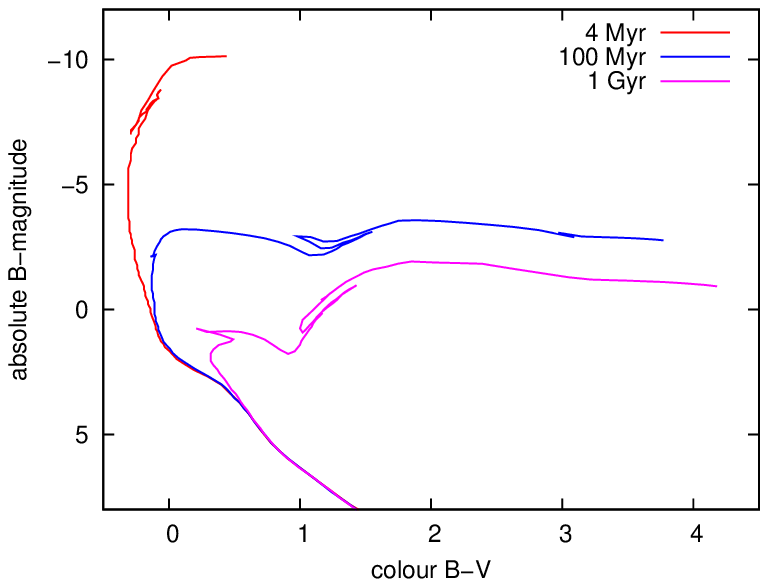}
\label{app:gen1}
\caption{Solar metallicity isochrones from the Padova group for three
  different ages of 4 Myr, 100 Myr and 1 Gyr}
\end{figure}

Fig. \ref{app:gen1} shows solar-metallicity isochrones for three different
ages of 4 Myr, 100 Myr and 1 Gyr as a colour-magnitude diagram. For very young
ages the main sequence reaches up to high masses and hence very high
luminosities $\rm\approx -10\,mag$. For the later stages the red giant branch
(RGB) at $\rm (B-V)\approx 1\ldots1.4$ is clearly visible and also the
asymptotic giant branch (AGB). Note that stars on the AGB can reach extremely
red colours of $\rm (B-V)\approx 4\,mag$ and at the same time get very bright
in the NIR ($\rm M_K\sim -10\,mag$) during the thermal pulsation (TP) phase.

\subsubsection*{Step 2: Select Initial Mass Function}
\begin{figure}
\includegraphics[width=\columnwidth]{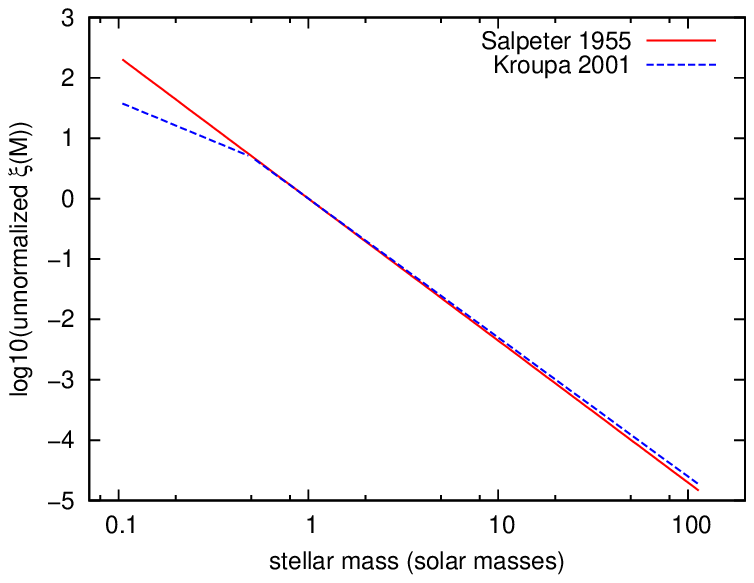}
\caption{Two examples of initial mass functions, here from
  \citet{Kroupa01} and \citet{Salpeter55}}
\label{app:gen2}
\end{figure}

In a second step we have to choose a parameterization for the number of stars as
function of their mass. Fig. \ref{app:gen2} shows two common IMFs, the Salpeter
and Kroupa IMFs. For masses $\ge 1\msun$ both predict roughly comparable number
of stars, but they differ in the low mass regime $\le 1\msun$. This will not
only affect mass-to-light ratios, but also have an impact on the resulting
spectra and in particular the chemical enrichment, since fewer low-mass stars
also means that less mass gets locked up in long-lived stars.

\subsubsection*{Step 3: Populate isochrones with stars}
\begin{figure}
\includegraphics[width=\columnwidth]{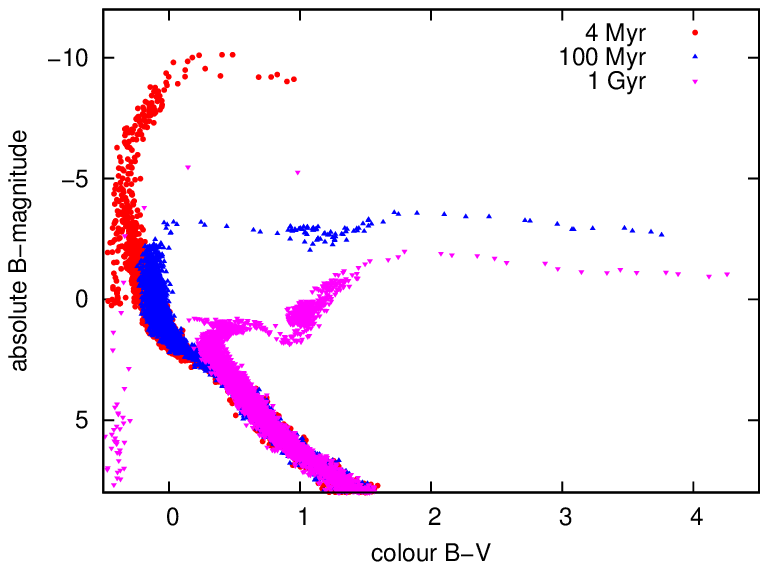}
\caption{Synthetic colour magnitude diagram where each point from the
  isochrone is assigned a numbers of stars based on the IMF. We added some
  scatter around each point to produce a smoother appearance.}
\label{app:gen3}
\end{figure}

With both isochrones and IMF at hand we can now populate the isochrones with
stars taken from an IMF. For each mass point on the isochrone the IMF tells us
how many stars were being formed with this mass. The results are shown in
Fig. \ref{app:gen3}, where we also added some scatter around the individual
points to give the artificial colour-magnitude diagram a smoother and more
realistic appearance. We also added at least one star to each point to bring
out all phases of stellar evolution. However, in reality, this might not be
the case, since some phases, e.g for very young and extremely hot white dwarfs
last for a very short timescale only.

\subsubsection*{Step 4: Assign a spectrum to each star}
\begin{figure}
\includegraphics[width=\columnwidth]{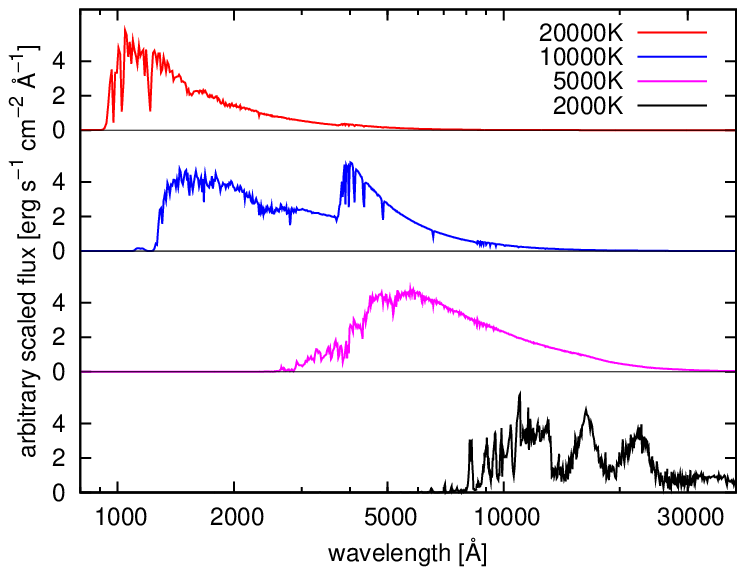}
\caption{Sample spectra from the Lejeune stellar library for stars with
  $\rm log g=4.0$ and $\rm T_{eff}=(20000, 10000, 5000, 2000)K$.}
\label{app:gen4}
\end{figure}

Next we assign a spectrum to each star in the colour-magnitude diagram. The
parameters required for this are either directly given in the isochrones
(e.g. the effective temperature $\rm T_{eff}$) or can be derived from other
parameters (e.g. the surface gravity $\rm \log g$ can be calculated from the
stellar mass, its bolometric luminosity and effective temperature). In
Fig. \ref{app:gen4} we show four sample spectra for stars with identical
$\log g=4.0$, but temperatures ranging from $\rm 20000\,K$ down to
$\rm 2000\,K$. Hot stars have a smooth slope and are bright in the
UV. Stars with $\rm T_{eff}\approx 10000\,K$ have prominent Balmer absorption
lines characteristic for spectral type A stars. Cool stars are dominated
by broad molecular absorption bands and only emit at long wavelengths. Not all
stellar parameters are directly covered by the library, all other values have
to be interpolated between neighbouring points in the library. Is is therefore
crucial for the stellar library to cover the full parameter range.

\subsubsection*{Step 5: Integrate spectra of all stars}
\begin{figure}
\includegraphics[width=\columnwidth]{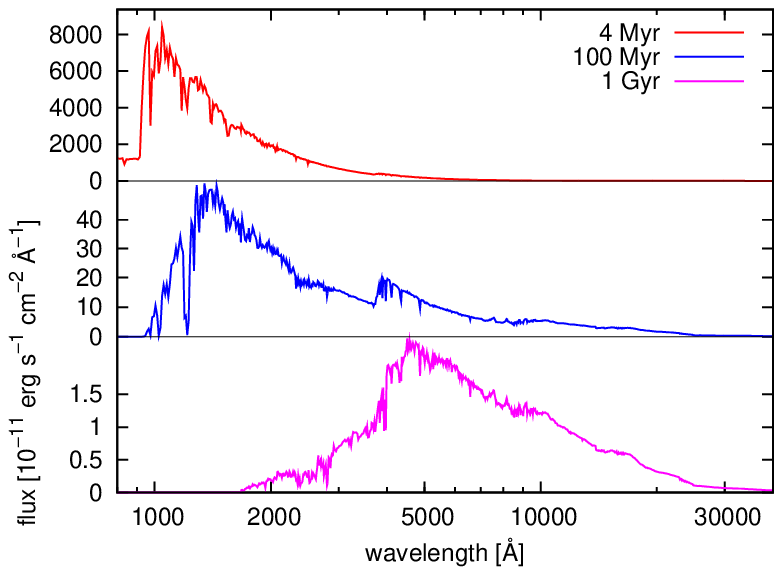}
\caption{Integrated starlight for the three isochrones shown in Fig.
  \ref{app:gen1}.}
\label{app:gen5}
\end{figure}

To create stellar isochrone spectra we now integrate the light of all stars.
Special care has to be taken to account for the different luminosities of
stars at their respective stages of evolution. In Fig. \ref{app:gen5} we show
spectra for the same ages discussed above, 4 Myr, 100 Myr, and 1 Gyr. The
youngest spectrum is almost completely dominated by the hottest and brightest
high-mass stars that also dominate part of the 100 Myr spectrum. For this
spectrum the highest-mass stars have already exploded as supernovae and with
them the flux in the FUV already has decreased considerably. In the optical
the $\rm 4000\AA$ break starts to appear, becoming stronger with time. After 1
Gyr there is little UV flux remaining, the spectrum is now dominated by
stars with lower masses $\approx 2\msun$.

\subsubsection*{Step 6: Compute gaseous line and continuum emission}
\begin{figure}
\includegraphics[width=\columnwidth]{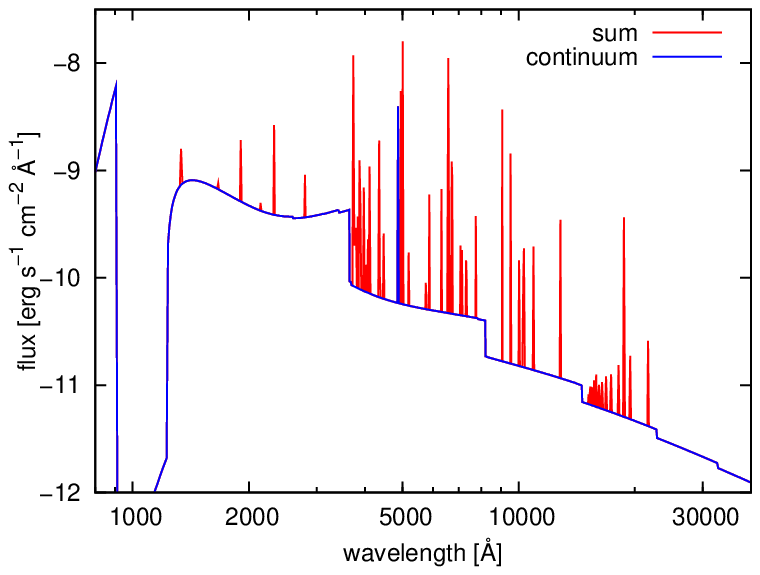}
\caption{Gaseous emission spectrum for solar metallicity. Shown in blue is the
continuum emission, and continuum and line emission in red.}
\label{app:gen6}
\end{figure}

Until now we have only dealt with stellar light. However, as mentioned earlier,
at young ages isochrones contain large contributions from high mass stars. These
emit a large fraction of their light in the UV and also produce significant
numbers of photons energetic enough to ionize hydrogen.  The energy injected
into the surrounding interstellar medium produces both emission lines and also
continuum emission. The detailed mechanisms of this are discussed in
Sect. \ref{subsec:input:lines} and shown in Fig. \ref{app:gen6}. Note that while
the strength of the heavy element emission lines depends on the metallicity of
the gas, the strength of the continuum emission and hydrogen emission lines
purely depends on the ionizing photon flux.

\subsubsection*{Step 7: Add gas emission to stellar spectrum}
\begin{figure}
\includegraphics[width=\columnwidth]{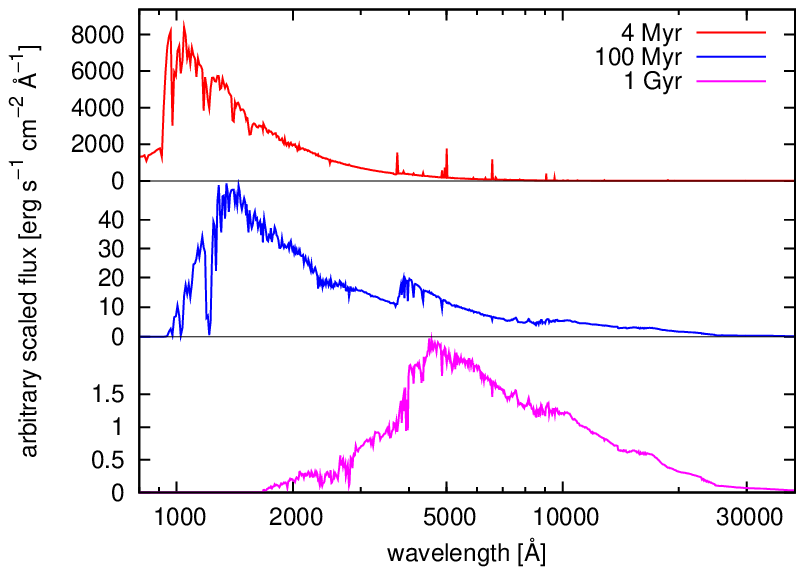}
\caption{Integrated isochrone spectra for the same isochrones shown in
  Figs. \ref{app:gen1} and \ref{app:gen5}, including both stellar and gaseous
  emission.}
\label{app:gen7}
\end{figure}

For each isochrone we now compute the gaseous emission spectrum based on the
metallicity and ionizing photon flux derived from isochrones and IMF. Since
the latter heavily depends on the age of the isochrone so does the strength of
the gaseous emission. In Fig. \ref{app:gen7} we show the same spectra as in
Fig \ref{app:gen5}, but now with gaseous emission included. For the first
spectrum at an age of $\rm 4\,Myr$ the emission lines clearly stand out in the
optical. However, for the later stages there are no changes compared to the
purely stellar spectra, since only stars with masses $\ga 20\msun$ and hence
very short lifetimes $\rm\la 10\,Myr$ produce the majority of the ionizing
flux.

In \galev those previous steps are repeated for all metallicities and ages
available from the isochrone set, yielding a full set of isochrone spectra
including both stellar and gaseous emission.

\subsection{Additional steps for star clusters}
Continuing from the library of isochrone spectra \galev can now compute the
finer grid required for the study of star clusters. 

\subsubsection*{Step 8: Interpolate between ages given by the
  isochrones}
\begin{figure}
\includegraphics[width=\columnwidth]{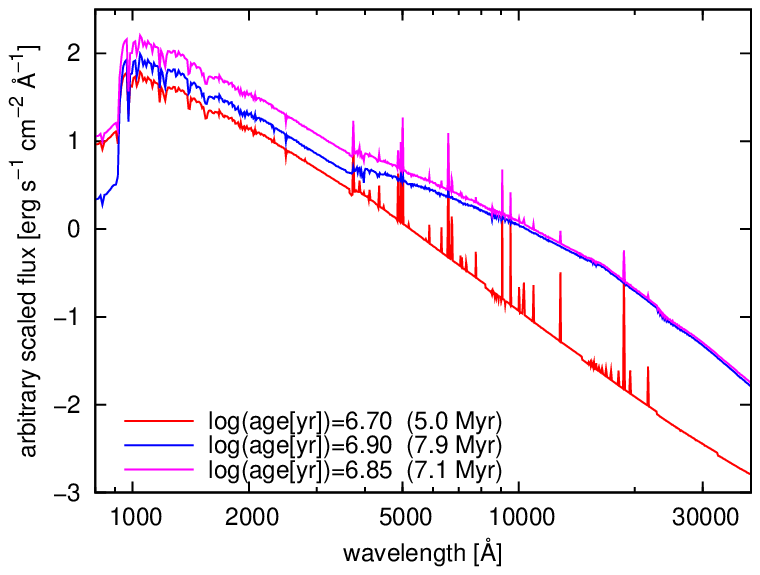}
\caption{Example for contributing isochrone spectra during interpolation in
  age.}
\label{app:sc1}
\end{figure}

The time resolution offered by the isochrones in general only spans a
relatively coarse grid with logarithmic age spacing. However, it is
advantageous to create a grid with smaller and linear age steps. This can be
done by interpolating additional ages to fill the gaps in the isochrone
grid. Since spectra vary roughly linearly with logarithmic time, i.e. changes
are larger at small ages and small at large ages we use this as basis for our
algorithm. In Fig. \ref{app:sc1} we show two isochrone spectra for ages of
$\log(t)=6.7$ and $\log(t)=6.9$ and the resulting spectrum for $\log(t) =
6.85$. The resulting spectrum was computed by $\rm Spec(log(t)=6.85) =
0.25\times(Spec(log(t)=6.7) + 3\times Spec(\log(t)=6.9)$.

To obtain a full grid to be used e.g. for age-dating star clusters one has to
repeat this procedure for each required time step and each metallicity.
The steps involved with applying dust extinction and computation of magnitudes
from spectra is explained in Steps 11 and 14 below.

\subsection{Additional steps for galaxies}
Galaxies are completely different from star clusters since they contain
multiple stellar populations, generally covering a range of ages and
metallicities. We explain the basic steps dealing with multiple populations
based on a toy-model of a galaxy made from only two populations with
different ages and metallicities. The subsequent explanations use the model of
an undisturbed elliptical galaxy.

\subsubsection*{Step 8: Compute initial abundances for next generation}
To derive the chemical evolution of galaxies we have to know four things: 1) The
total mass of the galaxy, including both the mass of stars and gas. 2) The star
formation history (SFH), i.e. how many stars are formed at each particular
time. 3) The life times of the stars formed. This in combination with the SFH
yields some form of star death record. And we have to know 4) the end products
of each star, i.e. the mass of its remnant, and the mass of gas and heavy
elements returned to the surrounding ISM. For each time we then have to keep
track of the masses of stars, gas and metals in the gas. All those quantities
are changed by star formation and the return of gas and heavy elements from
dying stars. The ratio of metals to gas is the crucial factor since it
determines both the spectra and lifetimes of the newly formed stars.

\subsubsection*{Step 9: Interpolate between ages AND metallicities}
For all galaxies with an extended SFH we will have stellar populations not
agreeing with the coarse grid given by the isochrones. We therefore have to
interpolate between isochrone spectra of different ages and also different
metallicities. The details of this process are described in the context of
star clusters above and are also shown in Fig. \ref{app:sc1}.

\subsubsection*{Step 10: Add up spectra weighted with SFH}
\begin{figure}
\includegraphics[width=\columnwidth]{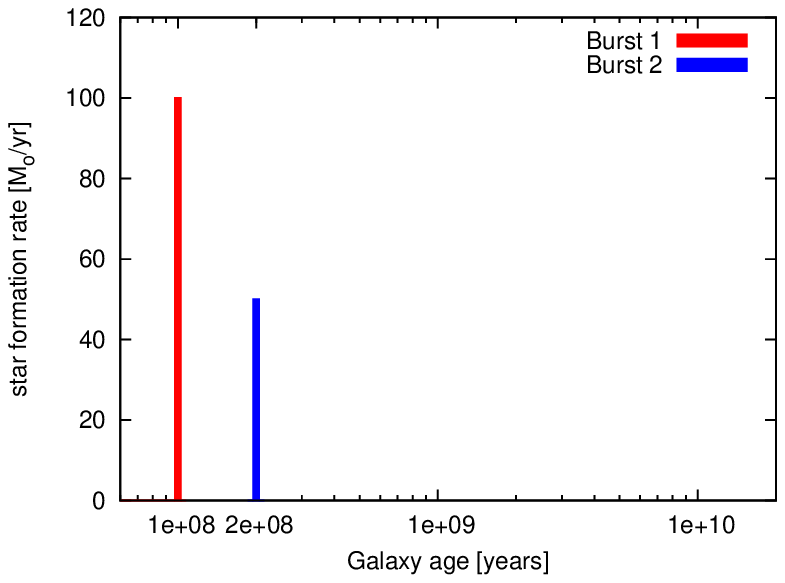}
\caption{For our toy-model we construct a very primitive star formation
  history of 2 bursts, with SFRs of 100$\,M_{\odot}\,yr^{1}$ and
  50$\,M_{\odot}\,yr^{1}$, each lasting $10^6$ years.}
\label{app:gal1}
\end{figure}

To ease the understanding how \galev assembles a galaxy spectrum from the
individual spectra of each of its constituent populations, we first consider a
toy model of a galaxy made from only two populations. Both populations are
described by intervals of $\rm 1\,Myr$ duration each, occuring at an age of
100 and 200 Myr and forming stars at a rate of $\rm 100\,\msun\,yr^{-1}$ and
$\rm 50\,\msun\,yr^{-1}$, respectively (see Fig. \ref{app:gal1}). Both
intervals are short compared to the age of the youngest isochrone so that they
each can be described as a population of one age. We further assume that the
earlier population (Burst 1) is formed with a metallicity of $\rm [Fe/H]=-1.7$
or $1/50 Z_{\odot}$, and the second (Burst 2) with a metallicity of $\rm
[Fe/H]=-0.7$ or $1/5 Z_{\odot}$.

\begin{figure}
\includegraphics[width=\columnwidth]{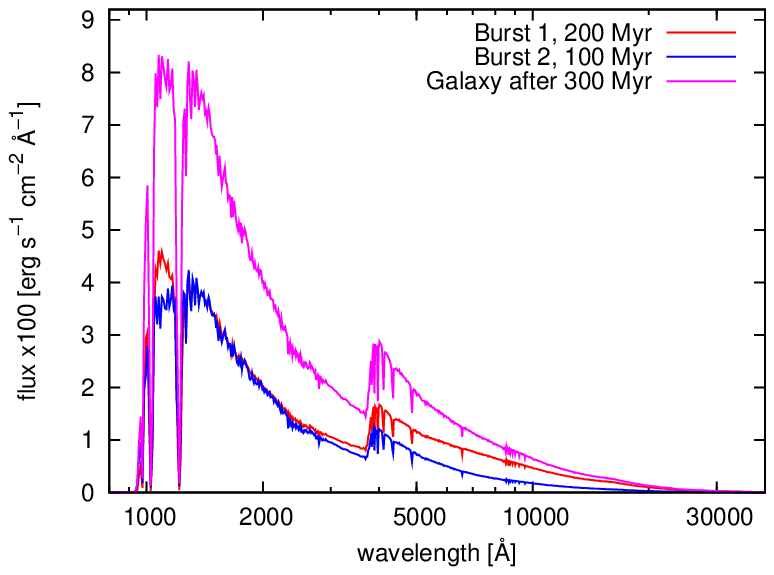}
\includegraphics[width=\columnwidth]{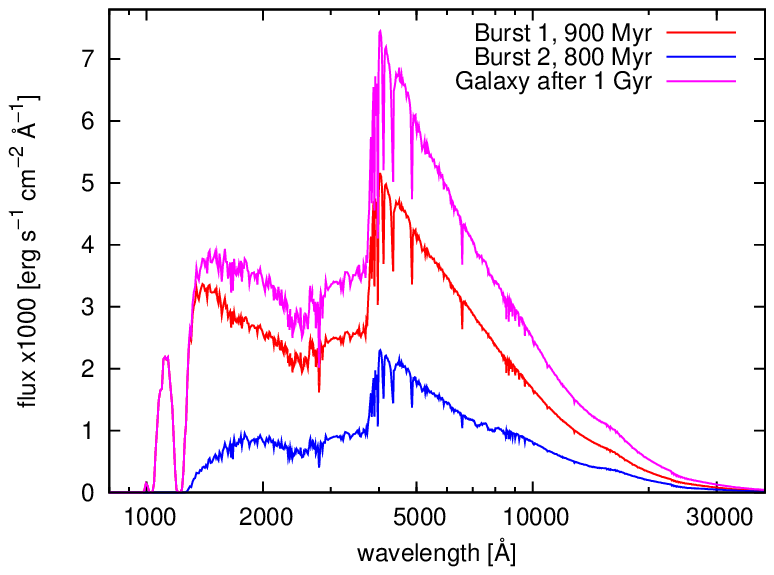}
\caption{Interpolated isochrone spectra for each burst and the resulting
  galaxy spectrum, for galaxy ages of 300 Myr (top panel) and 1 Gyr (bottom
  panel). }
\label{app:gal2}
\end{figure}

We will now show how to derive the spectrum of our toy-galaxy at two ages of
300 Myr and 1 Gyr. For each timestep t one needs to derive the ages $\tau$ and
metallicities of all populations formed prior to this time. Those are then
weighted by the SFR at their respective formation time $\rm SFR(t-\tau)$
multiplied with the length of a time step. Those are then added up to yield
the galaxy spectrum.
For our toy galaxy at an age of 300 Myr we know that Burst 1 has metallicity
$\rm [Fe/H]=-1.7$ and an age of 200 Myr. Burst 2 has metallicity $\rm
[Fe/H]=-0.7$ and an age of 100 Myr. The resulting spectrum thus can be
constructed by added up those two isochrone spectra.
The resulting spectrum is shown in the upper panel of Fig. \ref{app:gal2}.

The spectra for all other times are created in a similar way. Metallicities for
each population stay the same, while ages increase with time. The resulting
spectrum of our toy-model galaxy at an age of $\rm 1\,Gyr$ is shown in the
bottom panel of Fig. \ref{app:gal2}

In the following we will leave our toy-model galaxy and instead show the
remaining steps that are necessary to derive magnitudes for an elliptical
galaxy with a small amount of dust at redshift $\rm z=3$.

\subsubsection*{Step 11: Apply evolutionary correction}
\begin{figure}
\includegraphics[width=\columnwidth]{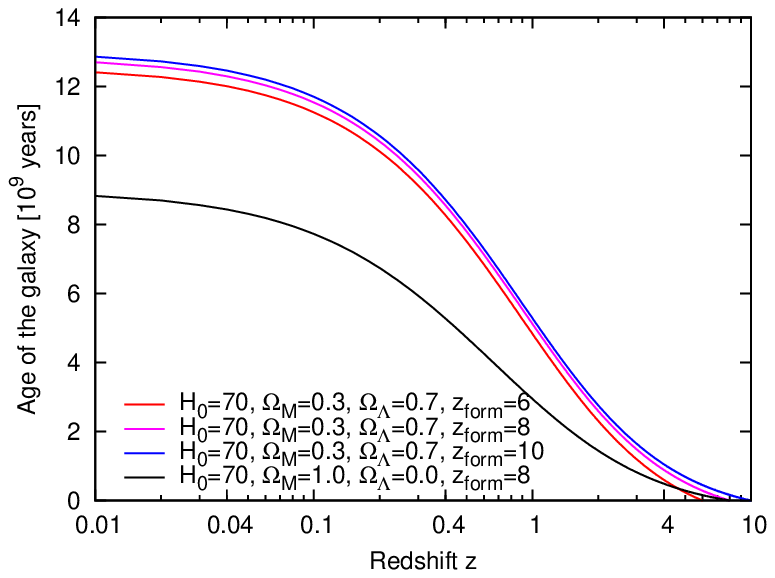}
\caption{Relation between galaxy age and redshift for different cosmological
  parameters and in particular varying formation redshifts.}
\label{app:gal4}
\end{figure}

Since we want to model the galaxy at a cosmologically significant redshift of
$\rm z=3$ we have to take evolutionary corrections into account, i.e. we
observe the galaxy at an earlier evolutionary state. We therefore have to know
the age of the galaxy at this redshift. In Fig. \ref{app:gal4} we show the
redshift-galaxy age relation for a small range of cosmological parameters. It
is obvious that the density parameters $\Omega_M$ and $\Omega_{\Lambda}$
influence the solution, but also the formation redshift $\rm z_{form}$, the
redshift at which the galaxy started forming stars. The impact of those
evolutionary corrections can be seen in Fig. \ref{app:galspecs}. The first row
shows the galaxy at redshift z=0 with an age of $\approx\rm 13\,Gyr$. The
second row shows the galaxy at its evolutionary state at $\rm z=3$; the galaxy
at redshift 3 started forming stars only $\rm\approx 1.5\,Gyr$ earlier.

\subsubsection*{Step 12: Apply extinction}
\begin{figure}
\includegraphics[width=\columnwidth]{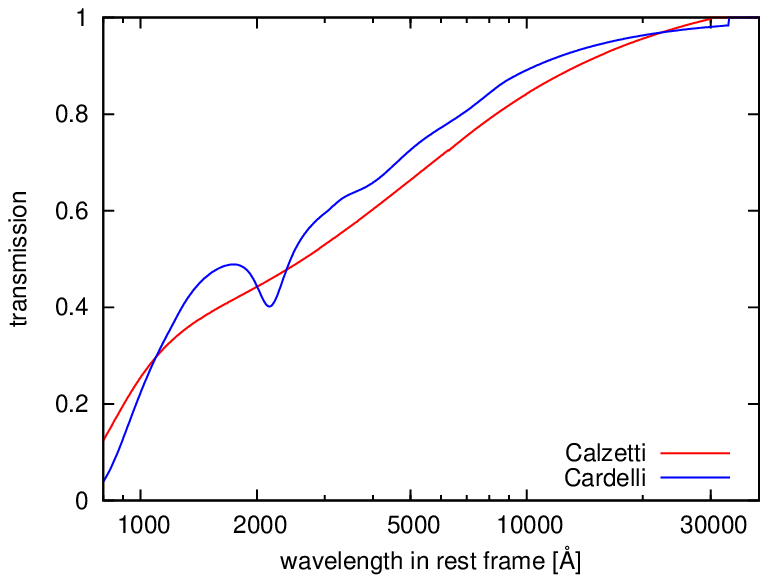}
\caption{Extinction curves from Calzetti and Cardelli. Both show very little
  transmission at short wavelengths indicating that most of the light gets
  absorbed.}
\label{app:gal3}
\end{figure}

In a next step we apply the attenuation due to interstellar dust. For our
example we choose the \cite{Calzetti+94} extinction law (see Fig.
\ref{app:gal3} and choose an intermediate degree of extinction, E(B-V)=0.2.
For comparison we also show the dust attenuation curve of \cite{Cardelli+89}.
Both extinction curves have in common that the transmission, i.e. the fraction
of light that remains unabsorbed, drops towards shorter wavelengths, leading
to a reddening of the galaxy light. The results on the spectrum can be seen
from the difference between the second and third row in Fig. \ref{app:galspecs}

\subsubsection*{Step 13: Redshift spectrum}
The spectrum now has to redshifted by a factor (1+z). Note that also the flux
has to reduced by the same factor (1+z) to properly account for the cosmic
expansion.

\subsubsection*{Step 14: Apply intergalactic attenuation curve}
\begin{figure}
\includegraphics[width=\columnwidth]{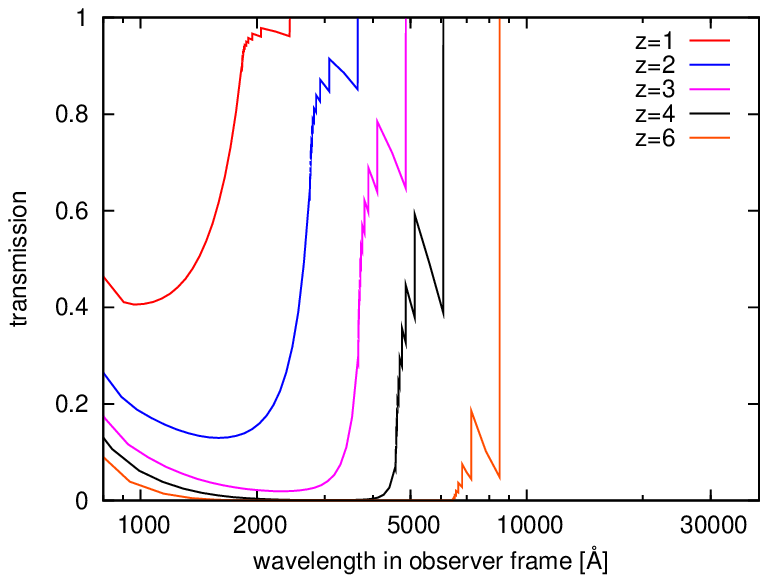}
\caption{Transmission of the intergalactic medium as function of observed
  frame wavelength for sources at different redshifts.}
\label{app:gal5}
\end{figure}

\begin{figure}
\includegraphics[width=\columnwidth]{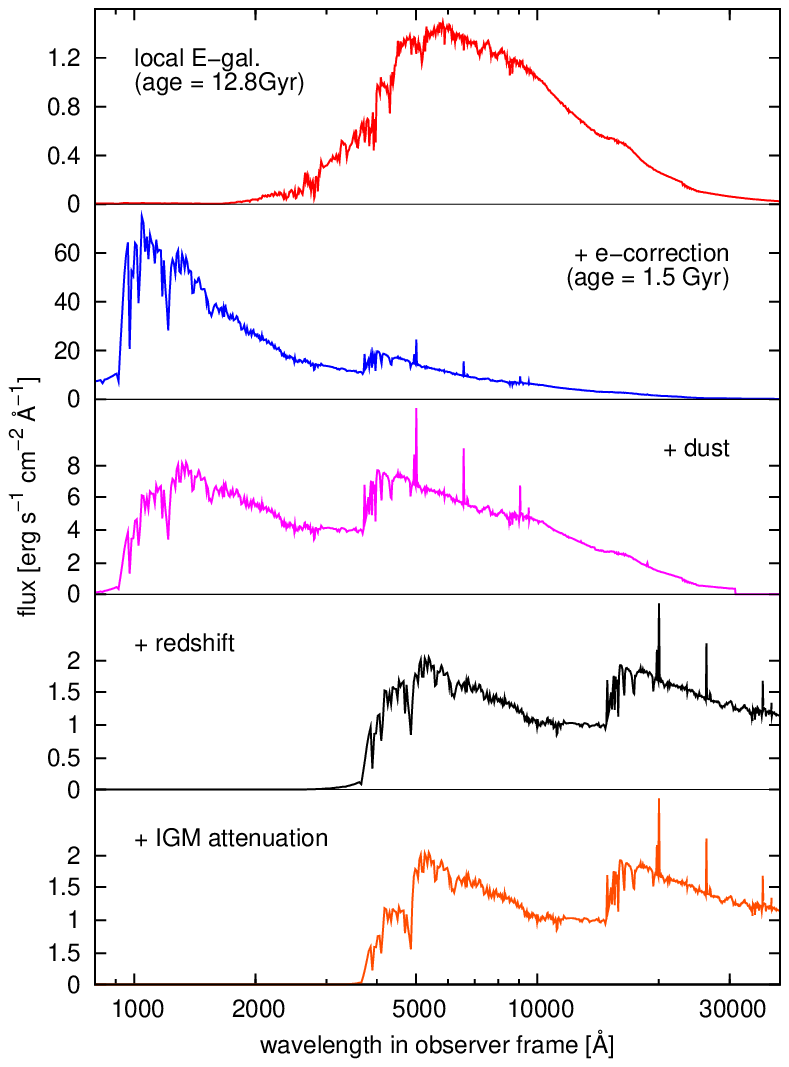}
\caption{Galaxy spectrum of an elliptical galaxy at different stages of the
  modelling process. Each following spectrum includes all effects shown above.}
\label{app:galspecs}
\end{figure}

Due to the high redshift at which we observe our galaxy we have to correct for
intergalactic absorption due to intervening neutral hydrogen clouds. These
absorb light shortwards of the Lyman-$\alpha$ line (1216 \AA) and hence reduce
the flux in this region. Fig. \ref{app:gal5} shows the transmission of the IGM
as function of observed frame wavelength for sources at different redshifts
$z=1\ldots 6$. For a galaxy at redshift 3 this means that $\approx 30\%$ of
the light between the Lyman-break (912 \AA) and the Lyman-$\alpha$ line is
absorbed, while shortwards of the Lyman-break essential all flux is absorbed
(hence the name Lyman-break). Fig. \ref{app:galspecs} shows in the two bottom
rows the spectrum with (5th row) and without (4th row) this attenuation.

All the previously mentioned effects are summarized in Fig.
\ref{app:galspecs}. The first row shows the spectrum of the galaxy at redshift
$z=0$ at an age of 13 Gyr. The second row still is at $\rm z=0$, but at an
evolutionary state already corresponding to $z=3$. The following panel shows
the spectrum with a reddening of $\rm E(B-V)=0.2\,mag$; Here most of the
far-UV flux is already absorbed by dust. The fourth row shows the spectrum
redshifted to $\rm z=3$ and the last row also takes intergalactic attenuation
into account.

\subsubsection*{Step 15: Convolve with filter response curve to compute
  magnitudes}
\begin{figure}
\includegraphics[width=\columnwidth]{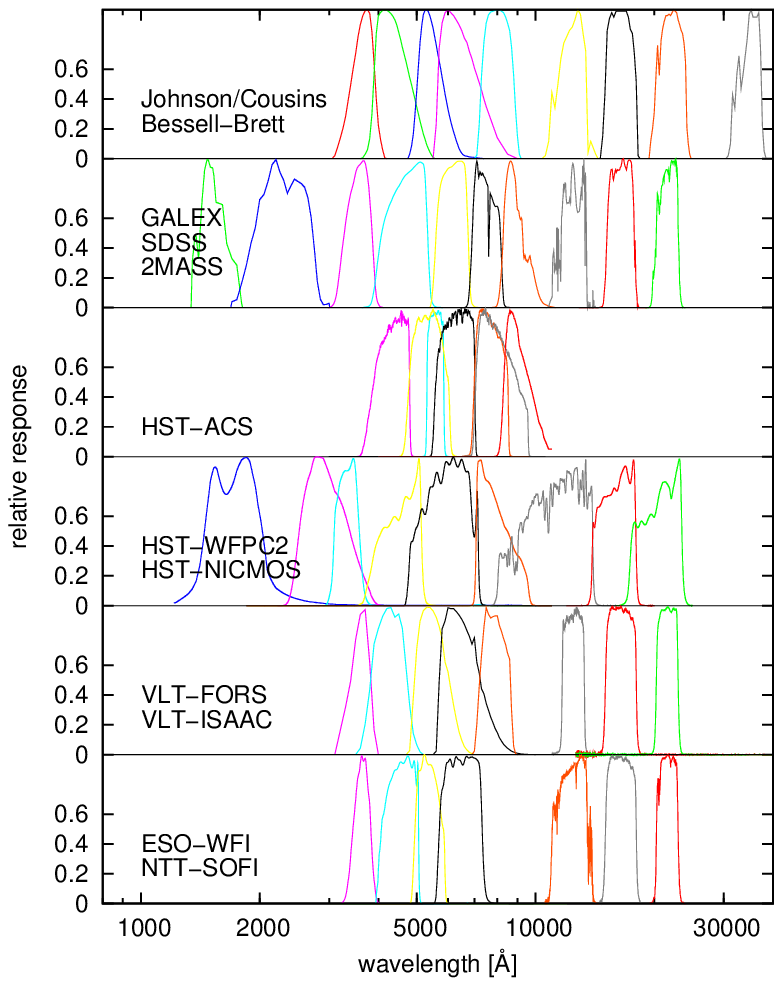}
\caption{Selection of the filters currently offered by \galev, ranging from the
far-UV to near-infrared.}
\label{app:gal6}
\end{figure}

We can now convolve the final spectrum with filter response curves. Therefore
each point in the spectrum is weighted with the relative filter response at
the corresponding wavelengths and then integrated over all wavelengths. To
convert the resulting fluxes into observable magnitudes we have to apply
zeropoints according to the requested magnitude system, e.g. Vega or AB.
In Fig. \ref{app:gal6} we show a wide selection of currently available
filters from different space and ground-based telescopes.

\subsubsection*{Step 16: Add distance modulus}
\begin{figure}
\includegraphics[width=\columnwidth]{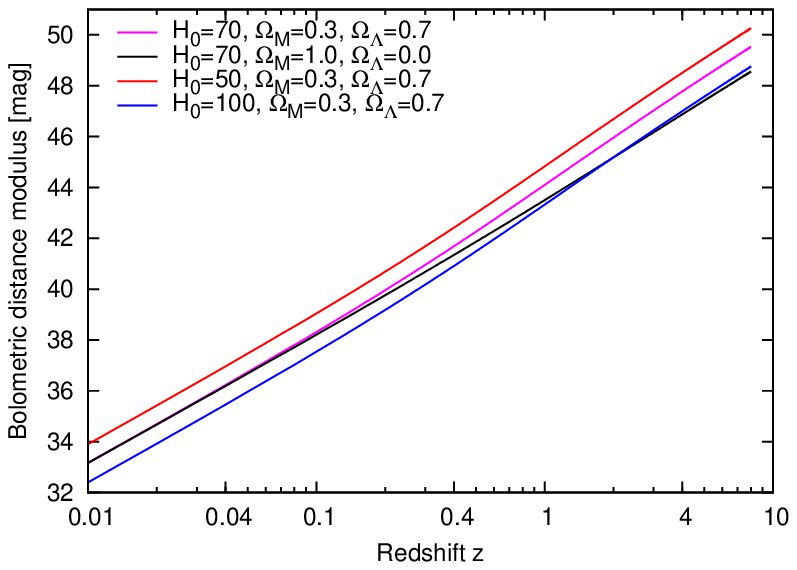}
\caption{Bolometric distance modulus as function of redshift for a range of
  cosmological parameter sets.}
\label{app:gal7}
\end{figure}

In a very last step we can convert the absolute magnitudes obtained with the
filters into apparent magnitudes. This is done by simply adding the bolometric
distance modulus for the particular redshift. In Fig. \ref{app:gal7} we show
the evolution of the distance modulus with redshift for a small selection of
possible cosmological parameter sets. For the previously studied galaxy at
redshift $\rm z=3$ we have to add a distance modulus of 47 mag.

\bsp
\bibliography{galev}
\bibliographystyle{aa}

\label{lastpage}

\end{document}